\documentclass[11pt,a4paper]{article}

\usepackage[top=3cm, bottom=4cm, left=3cm, right=3cm]{geometry}

\usepackage{amsmath,amssymb}
\usepackage{amsthm}
\usepackage[active]{srcltx}
\usepackage{cite}
\usepackage{graphicx}
\usepackage{psfrag}

\newcommand{\emi}{({\em i}\,) }
\newcommand{\emii}{({\em ii}\,) }
\newcommand{\emiii}{({\em iii}\,) }

\newcommand{\R}{\mathbb{R}}

\newcommand{\z}{{\mathbf{z}}}
\renewcommand{\r}{{\mathbf{r}}}
\newcommand{\Lc}{{\cal L}}
\newcommand{\Mc}{{\cal M}}
\newcommand{\Oc}{{\cal O}}
\newcommand{\Uc}{{\cal U}}
\newcommand{\Wc}{{\cal W}}
\newcommand{\Rc}{{\cal R}}

\newcounter{mnotecount}[section]

\newtheorem{thr}{Theorem}

\numberwithin{equation}{section}

\begin{document}

\title{Does time always slow down as gravity increases?}
\author{Andrzej Oko{\l}\'ow }
\date{February 24, 2021}

\maketitle
\begin{center}
{\it  Institute of Theoretical Physics, Warsaw University\\ ul. Pasteura 5, 02-093 Warsaw, Poland\smallskip\\
oko@fuw.edu.pl}
\end{center}
\medskip

\begin{abstract}
We consider gravitational time dilation between stationary observers and present examples, which contradict the statement that ``time slows down as gravity increases''. {We show furthermore that this statement cannot be true in general, if strength of gravity is defined in an observer independent manner. We provide also a pedagogical introduction to gravitational time dilation, and} discuss aspects of {this phenomenon}, which are often omitted in textbooks on general relativity. 
\end{abstract}\medskip

\begin{center}
\fbox{\begin{minipage}{0.9\textwidth}
\small This is the Accepted Manuscript version of an article accepted for publication in European Journal of Physics, including the subsequent correction of some linguistic errors. IOP Publishing Ltd is not responsible for any errors or omissions in this version of the manuscript or any version derived from it. The Version of Record is available online at https://doi.org/10.1088/1361-6404/ab60bb.
\end{minipage}}
\end{center}\medskip

\section{Introduction}

The fact that general relativity (GR) predicts a phenomenon called gravitational time dilation, is well-known not only among physicists, but also among amateurs interested in GR. There is the notion that the time dilation ratio is correlated with strength of gravity, and it seems that this notion is widespread to some extent---the statement ``time slows down as gravity increases'' appears, e.g., in  \cite{ripples,inf-tech}, claims that the stronger the gravitational field, the slower time goes, that clocks close to massive bodies run more slowly and similar, can be found on web pages treating of physics or GR, and in presentations available via the Internet.   

{One of the two goals} of the paper, is to check if the statement ``time slows down as gravity increases'' is correct in general. To the best of our knowledge, there is no strict proof of this statement---it seems rather that it is based on some particular examples, like time dilation between static observers in the Schwarzschild spacetime, or time dilation in a weak gravitational field, which are often discussed in textbooks on GR. Therefore our strategy will be to try to disprove the statement by providing appropriate counterexamples, rather than to prove it. A simple description of gravitational time dilation is available in the case of stationary observers, and so while looking for the counterexamples, we will restrict ourselves exclusively to this type of observers. We will show that it is not difficult to find such counterexamples even in the Schwarzschild spacetime.

{Except presenting the counterexamples, we will prove one general result, concerning the relation between gravitational time dilation and strength of gravity. To make this relation precise, one has to define, what strength of gravity is. A natural way to do this, is to choose appropriately a scalar field derived from the spacetime metric, and treat the value of the scalar field as a measure of the strength. We will prove that there does not exist any scalar field, derived from the metric in an observer independent way, such that time always ``slows down'' as the value of the field increases.}

In many textbooks on GR, gravitational time dilation is presented rather cursorily. The other goal of this paper is {to provide a pedagogical introduction to this phenomenon and to discuss those its} aspects, for which there is usually no room in both introductory and advanced courses on GR---taking into account the ``fame'' of time dilation mentioned above, this phenomenon certainly deserves a more complete and thorough treatment. 

First of all, we would like to enhance the geometric approach to gravitational time dilation. The main motivation for this is the fact that expressions like ``time slows down'', used often to describe time dilation, are actually simplifications, which can be a source of some misconceptions. The geometric approach appropriately emphasizes the aspects of this phenomenon omitted by such simplifications. 

Furthermore, we will show that in general there are some ambiguities related to time dilation, and that these ambiguities disappear, once we restrict ourselves to time dilation perceived by stationary observers. We will also present an interpretation of such observers as a generalization of a rigid body in Newtonian mechanics. Moreover, we will discuss relationships between, on one hand, gravitational time dilation and gravitational redshift and, on the other hand, spacetime curvature and the observer's acceleration---we will conclude that both phenomena are not closely related to the curvature, but are conditioned by the acceleration.

In the appendix we will present a suggestion, based on the geometric approach mentioned above: How to teach gravitational time dilation by means of a polar coordinate system on the Euclidean plane.

\section{Preliminaries}

The statement ``time slows down as gravity increases'' and others, are rather simplified and therefore imprecise descriptions of gravitational time dilation because: 
\begin{enumerate}
\item the lapse of time is not an observable in GR,
\item there is not a unique measure of gravitational field strength in GR.
\end{enumerate} 

The lapse of time itself is not an observable, but {\em the lapse of time along a timelike curve $\kappa$ from an (earlier) event $\z_1\in \kappa$, to a (later) event $\z_2\in\kappa$}, can both be measured and calculated. In other words, the lapse of time along an interval of a timelike curve, is an observable in GR.

On the other hand, the gravitational field in GR is described by a Lorentzian metric, and many different scalar fields, such as the Ricci scalar curvature, can be derived form the metric. It is rather difficult to select one of these scalar fields, argue {that its value is the best measure of strength of the gravitational field, and discard the other fields as useless.}

Thus to describe gravitational time dilation in a precise way, one has to compare the lapses of time along intervals of different timelike curves. Furthermore, trying to find a relation between the ratio of time dilation and strength of gravity, one has to choose a particular scalar field derived from the metric, as a measure of this strength.

\subsection{Gravitational time dilation}

\subsubsection{Definition \label{desc}}

Let $(\Mc,g)$ be a spacetime, where $\Mc$ is a four-dimensional manifold, and $g$ a metric of signature $(-,+,+,+)$ defined on $\Mc$. From the point of view of geometry, the basic task of a metric is to define a scalar product between every pair of vectors, tangent to $\Mc$ at a given point. We will denote by $g(X,X')$ such a product between vectors $X,X'$, defined by the metric $g$.        

Suppose that $\kappa_{\z_1\z_2}$ is an interval of a timelike curve $\kappa\subset\Mc$, determined by its endpoints $\z_1$ and $\z_2$. Assume also that  $\kappa$ is the world line of an observer equipped with a clock, and the readings on the clock are $\tau_1$ at $\z_1$, and $\tau_2$ at $\z_2$. Then the measured lapse $\Delta \tau_m$ of proper time (or the measured proper time interval) along $\kappa_{\z_1\z_2}$, is the difference between $\tau_2$ and $\tau_1$:  
\[
\Delta \tau_m=\tau_2-\tau_1.
\]

On the other hand, if the map $\lambda\mapsto\kappa(\lambda)$ parameterizes the curve $\kappa$ in such a way that
\begin{align}
\z_1&=\kappa(\lambda_1),&\z_2&=\kappa(\lambda_2),
\label{z-lambda}
\end{align}
then one can calculate the lapse $\Delta \tau_c$ of proper time (or the proper time interval) along $\kappa_{\z_1\z_2}$  as follows:
\begin{equation}
\Delta\tau_c=\int_{\lambda_1}^{\lambda_2}\sqrt{|g(\dot{\kappa},\dot{\kappa})|}\,d\lambda,
\label{D-tau}
\end{equation}
where $\dot{\kappa}(\lambda_0)$ denotes the vector tangent to $\lambda\mapsto\kappa(\lambda)$ at $\lambda_0$.

Taking into account the similarity of the r.h.s. of \eqref{D-tau} to the expression for the length of a curve in the Euclidean space, we call $\Delta\tau_c$ the  ``spacetime length'' of the interval  $\kappa_{\z_1\z_2}$.      

Given an interval $\kappa_{\z_1\z_2}$, the values $\Delta\tau_c$ and $\Delta\tau_m$ may differ if the latter is measured by a clock in reality, where acceleration and tidal forces may be present, and influencing its functioning. Therefore throughout the paper, we will apply the ``clock hypothesis'' (see e.g. \cite{clock}), that is, we will assume that $\Delta\tau_m$ is measured by an ideal clock, and therefore it coincides with $\Delta\tau_c$. From here on, we will drop the indices $m$ and $c$, and denote the lapse of proper time simply by $\Delta\tau$. We will also omit the adjective ``proper'' and will call $\Delta\tau$  ``lapse of time'' or ``time interval''.   

Consider now two observers $\Oc$ and $\Oc'$, of world lines $\kappa$ and $\kappa'$ respectively, both equipped with ideal clocks. What is the time dilation between the two observers? To answer this question, we choose intervals of $\kappa$ and $\kappa'$, measure or calculate the lapses of time $\Delta \tau$ and $\Delta \tau'$  along the chosen intervals, and describe the gravitational time dilation by means of the ratio
\begin{equation}
D=\frac{\Delta\tau'}{\Delta\tau},
\label{dil}
\end{equation}
with time dilation being equivalent to $D\neq 1$. 

\begin{figure}
\psfrag{h}{$\kappa$}
\psfrag{h'}{$\kappa'$}
\psfrag{hz}{$\kappa_{\z_1\z_2}$}
\psfrag{hz'}{$\kappa'_{\z'_1\z'_2}$}
\psfrag{z1}{$\z_1$}
\psfrag{z2}{$\z_2$}
\psfrag{z1'}{$\z'_1$}
\psfrag{z2'}{$\z'_2$}
\begin{center}
\includegraphics{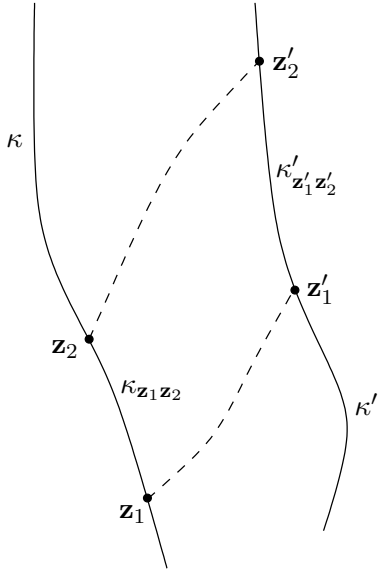}
\end{center}
\caption{Choice of intervals \label{dil-fig}}
\end{figure}

But {\em how should we choose the intervals?} It is obvious that the intervals cannot be chosen arbitrarily, because then the ratio \eqref{dil} can be arbitrary: for example, choosing a ``short'' interval of $\kappa$ and a ``long'' interval of $\kappa'$, compare the corresponding lapses of time and claim that ``the time of the observer $\Oc$ runs slower than the time of $\Oc'$''. Thus we have to use {\em a physical/geometrical criterion},  which would tell us how to choose intervals for the comparison. 

The standard method for choosing intervals (see \cite{haicek,grav-inert}) reads as follows: The observer $\Oc$ selects arbitrarily two events $\z_1$ and $\z_2$, both lying on the world line $\kappa$. At $\z_1$, $\Oc$ sends a light flash to the observer $\Oc'$, who receives it at an event $\z'_1$. At $\z_2$,  $\Oc$ again sends a light flash to $\Oc'$, who receives it at an event $\z'_2$. In geometrical terms, the events $\z'_1$ and $\z'_2$ are selected as intersection points between the world line $\kappa'$ and the null geodesics, starting at the events $\z_1$ and $\z_2$ respectively (see Figure \ref{dil-fig}, where the geodesics are depicted as dashed lines). The outcome of the standard method is a pair of intervals $\kappa_{\z_1\z_2}$ and $\kappa'_{\z'_1\z'_2}$, and the lapses of time along them are used to calculate the ratio $D$ in Equation \eqref{dil}. 

To summarize: from the geometric point of view, gravitational time dilation appears if ``spacetime lengths'' of appropriately paired intervals of two world lines, are distinct. 

\subsubsection{Remarks \label{rem}}
 
The notion of {\em gravitational} time dilation, as defined above, is very broad and includes some cases, where gravity seems to play no role (two observers in the Minkowski spacetime whose world lines are geodesics). However, the term ``gravitational time dilation'' is used in the literature even in such a broad context \cite{grav-inert}, and therefore we will continue to use it in this paper. 

Let us emphasize that gravitational time dilation is a different phenomenon than special relativity time dilation:  the former concerns proper time intervals of two arbitrary observers, while the latter relates a coordinate time interval in an inertial frame, and a proper time interval of an observer, moving with respect to the frame with constant velocity.

In the case of an arbitrary pair $\kappa$ and $\kappa'$ of timelike world lines,  in an arbitrary spacetime
\begin{enumerate}
\item the standard method for choosing an interval $\kappa'_{\z'_1\z'_2}\subset\kappa'$ for a fixed interval $\kappa_{\z_1\z_2}\subset\kappa$, can give ambiguous results, and this can give rise to an ambiguity of the ratio \eqref{dil};  
\item for a fixed dependence of  $\kappa'_{\z'_1\z'_2}$ on $\kappa_{\z_1\z_2}$, selected from those offered by the standard method, the ratio \eqref{dil} depends on both $\z_1$ and $\z_2$;
\item there exists a natural extension of the standard method, which seems to be as valid as the original, but with significantly more ambiguity.   
\end{enumerate}
Let us now justify the first and third statements, since the second one seems to be rather obvious.

\begin{figure}
\psfrag{t}{$t$}
\psfrag{z1}{$\z_1$}
\psfrag{z2}{$\z_2$}
\psfrag{z'1}{$\z'_1$}
\psfrag{z'2}{$\z'_2$}
\psfrag{zb'1}{$\bar{\z}'_1$}
\psfrag{zb'2}{$\bar{\z}'_2$}
\psfrag{c}{$\kappa$}
\psfrag{c'}{$\kappa'$}
\begin{center}
\includegraphics{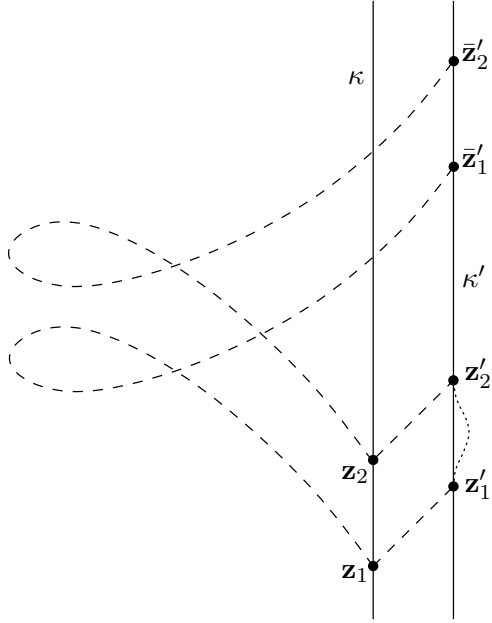}
\end{center}
\caption{Ambiguity of the standard method \label{geo}}
\end{figure}

\paragraph{Ambiguity of the standard method} Let us consider the Schwarzschild spacetime  with the metric \cite{schw}
\begin{equation}
g=-\Big(1-\frac{2M}{r}\Big)dt^2+\Big(1-\frac{2M}{r}\Big)^{-1}dr^2+r^2(d\theta^2+\sin^2\theta\,d\varphi^2), \quad r>0,
\label{schw}
\end{equation}
where $M$ is a positive constant, and let  $\Oc$ and $\Oc'$ be  {\em static}\footnote{An observer in the Schwarzschild spacetime is static if the coordinates $(r,\theta,\varphi)$ do not change along his world line. For a general definition of static observer see Section \ref{dfs}.} observers in this spacetime. Assume that \emi the values of the angular coordinates $(\theta,\varphi)$ of both observers are the same, \emii the value of the radial coordinate $r$ of $\Oc$, is equal to $3M$, which is the ``radius'' of a circular orbit of a light ray in this spacetime (see e.g. \cite{wald}) and \emiii the value of the radial coordinate of $\Oc'$, is greater than  $3M$.

It then turns out that the world line $\kappa$ of $\Oc$ can be connected with the world line $\kappa'$ of  $\Oc'$, by means of both a {\em radial} null geodesic $\gamma$ and a {\em non-radial} null geodesic $\bar{\gamma}$---the latter one begins at a point of $\kappa$, circles the world volume $r=3M$ receding from it, and meets the world line $\kappa'$. This allows for the association of the interval $\kappa_{\z_1\z_2}$ of $\kappa$, with two distinct intervals $\kappa'_{\z'_1\z'_2}$ and $\kappa'_{\bar{\z}'_1\bar{\z}'_2}$ of $\kappa'$,  as shown on Figure \ref{geo}. In this particular case, the lapse $\Delta\tau'$ of time along $\kappa'_{\z'_1\z'_2}$, is equal to the lapse $\Delta\bar{\tau}'$ along $\kappa'_{\bar{\z}'_1\bar{\z}'_2}$ (see Section \ref{disc}). However, slight deformations of the world line $\kappa'$ between $\z'_1$ and $\z'_2$ (the dotted line on Figure \ref{geo}), can be made so that $\Delta\tau'\neq\Delta\bar{\tau}'$. 

\paragraph{Extension of the standard method} In order to associate an interval  $\kappa'_{\z'_1\z'_2}$ with $\kappa_{\z_1\z_2}$, the observer $\Oc$ can send to $\Oc'$ a pair of {\em free massive particles}, instead of a pair of light flashes{---}in this case the resulting interval $\kappa'_{\z'_1\z'_2}$ would be chosen by means of {\em timelike} geodesics, instead of null geodesics.

In every spacetime, both timelike and null geodesics, do not depend directly on non-gra\-vi\-ta\-tio\-nal fields\, which exist in this spacetime, but are fully determined by the gravitational one. Therefore it seems that if null geodesics are good for pairing of intervals, when defining gravitational time dilation, then timelike ones should also be good for the same purpose.      

It is then rather natural to extend the standard method, by allowing observers to also use free massive particles for the pairing of intervals. The extended method is, however,  much more ambiguous than the original one (the reason is that, unlike light flashes, massive particles can be sent in the same direction with velocity of various values).

\paragraph{Summary} In the general case, in order to establish the value of the time dilation ratio \eqref{dil}, it is not enough to specify a pair of observers, but it is necessary to specify a pair of intervals of the observers' world lines. The pairing of the intervals is in general not unique for either the standard or extended method, and so this ambiguity affects the time dilation ratio. Consequently, the description of gravitational time dilation in the general case, is rather cumbersome. 

Therefore in the rest of the paper, we will restrict ourselves to a certain class of spacetimes, and a certain class of observers, where gravitational time dilation can be described in a simple and unambiguous way.

\subsection{Gravitational time dilation between stationary observers}

Considerations in the remaining part of this paper, will be based on the notion of Killing vector fields. In order to fit the presentation below to the geometric approach to time dilation, the introduction to these vector fields will emphasize some properties of their integral curves. An alternative simple introduction to this subject can be found in \cite{red-kill}. 

Let us fix some notation here. Given (local) coordinate system $(x^\alpha)_{\alpha=0,1,2,3}$ on a spacetime, the symbol $\partial_{x^\alpha}$ will denote a vector field such that all its components in the system are zero, except the $\alpha$-th component, which is equal $1$ e.g. $\partial_{x^2}=(0,0,1,0)$. The symbol $f_{,\alpha}$ will denote the partial derivative of a function $f$, with respect to the coordinate $x^\alpha$. Finally, given vector field $X$, we will denote the derivative of $f$ along $X$ by $Xf$. If $X=X^\alpha\partial_{x^\alpha}$, then $Xf=X^\alpha f_{,\alpha}$.            

\subsubsection{Flow of a vector field}
\begin{figure}
\psfrag{K}{$K$}
\psfrag{p}{$\phi_s$}
\psfrag{h}{$\kappa$}
\psfrag{z}{$\z$}
\psfrag{M}{$\Mc$}
\psfrag{l0}{$\lambda_0$}
\psfrag{los}{$\lambda_0+s$}
\psfrag{l}{$\lambda$}
\begin{center}
\includegraphics{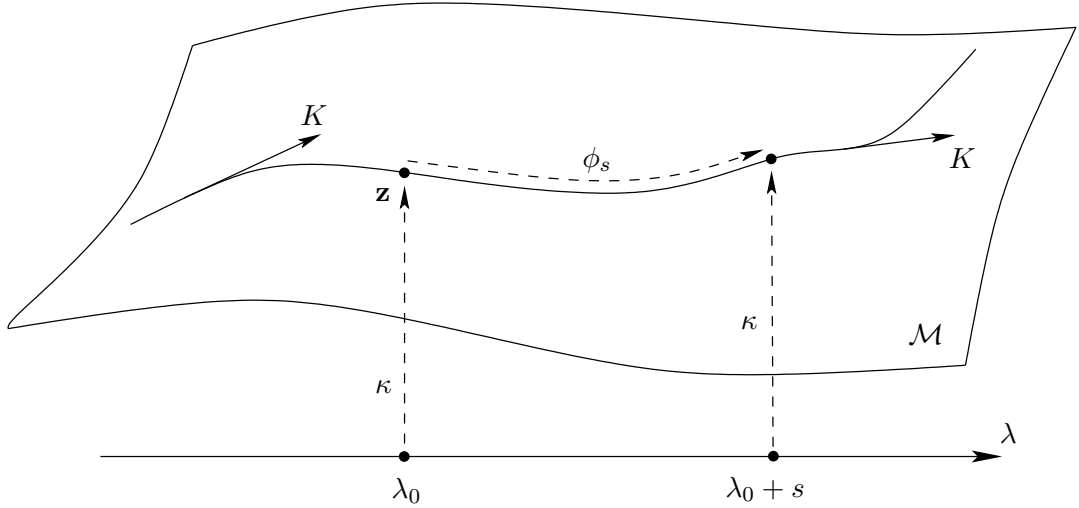}
\end{center}
\caption{Action of $\phi_s$ \label{shift}}
\end{figure}

Consider the set of all integral curves of a vector field $K$, defined on a subset $\Uc$ of a spacetime $\Mc$. Given number $s$, we denote by  $\phi_s$ a map on $\Uc$, which shifts points along the curves, by increasing by $s$ the parameter along each curve. Strictly speaking, if   
\[
\lambda\mapsto\kappa(\lambda)
\]   
is an integral curve of $K$ passing through the event $\z=\kappa(\lambda_0)$, then (see Figure \ref{shift})
\[
\phi_s(\z)=\kappa(\lambda_0+s).
\]
The set $\{\phi_s\}$ of all such maps (which differ by values of $s$), is called {\em flow of the vector field $K$}. 

Each map $\phi_s$  in the flow of $K$ can be used to shift tensor fields along the integral curves of $K$. If $f$ is a scalar field, then the value of the shifted scalar field $\phi^*_sf$ at a point $\z$, is equal to $f(\phi_s(\z))$.  

To shift a vector field $X$, we act by $\phi_s$ on every integral curve of $X$: if $\lambda\mapsto \chi(\lambda)$ is such a curve, then $\lambda\mapsto \phi_s(\chi(\lambda))$ is the shifted curve. The resulting set of all the shifted curves defines the shifted vector field, which is usually denoted by $\phi_{s*}X$.     

To shift a metric $g$ on $\Mc$, let us recall that complete information about $g$ can be encoded in its orthonormal frame, that is, in a set of four vector fields $(X_\alpha)_{\alpha=0,1,2,3}$ such that
\[
g(X_\alpha,X_\beta)=
\begin{cases}
-1 & \text{if $\alpha=\beta=0$,}\\
1 & \text{if $\alpha=\beta\neq 0$,}\\
0 & \text{if $\alpha\neq\beta$}.
\end{cases}
\]    
Thus to shift the metric $g$ along the integral curves of $K$, it is enough to shift its orthonormal frame $(X_\alpha)$. More precisely, we shift each vector field in the frame by means of $\phi_{-s}$ obtaining vector fields $(\phi_{-s*}X_\alpha)$, which constitute an orthonormal frame of the shifted metric $\phi^*_sg$.

\subsubsection{Killing vector fields }

Let $K$ be a vector field defined on a subset $\Uc$ of a spacetime $\Mc$, equipped with a metric $g$, and let $\{\phi_s\}$ be the flow of $K$. The vector field $K$ is called a {\em Killing vector field (of the metric $g$)} if the metric $g$ does not change, when shifted along the integral curves of $K$:
\begin{equation}
\phi_s^*g=g
\label{phisg-g}
\end{equation}
for every $\phi_s$ in the flow of $K$.

Let us now calculate the derivative of the equation above, with respect to $s$ at $s=0$. The derivative of the l.h.s. of \eqref{phisg-g}, is called the {\em Lie derivative} of the metric with respect to $K$, and is denoted by $\Lc_Kg$. The  r.h.s. of \eqref{phisg-g} does not depend on $s$, and therefore its derivative is zero. Thus we obtain from \eqref{phisg-g} the following equation 
\begin{equation}
\Lc_Kg=0,
\label{LK-g}
\end{equation}
which is an alternative definition of Killing vector field.

Suppose that a vector field $K$ is non-zero everywhere on an open set $\Uc$. Then $K$ is a Killing vector field on $\Uc$ if and only if on a neighborhood of every point in $\Uc$, there exists a coordinate system $(x^\alpha)_{\alpha=0,1,2,3}$ such that
\[
K=\partial_{x^0},
\]
and all the components $g_{\alpha\beta}$ of the metric in this system, do not depend on the coordinate $x^0$ (a method used to construct such coordinates is presented in Section \ref{rigid}).     

We advise the reader, who is unfamiliar with Killing vector fields, to study two simple examples of such fields: the first one presented in Appendix \ref{kill-ap}, and then the one given by Equation \eqref{K-u-acc}.

\subsubsection{Stationary and static observers \label{dfs}}

If a Killing vector field $K$ is timelike on a subset $\Wc\subset \Mc$, then $\Wc$ is said to be {\em stationary}. An observer $\Oc$ is called {\em stationary} if his (timelike) world line $\kappa$ is an integral curve of a Killing vector field\footnote{Strictly speaking, the world line of a stationary observer, is the {\em image} of an integral curve of a Killing vector field.}.

Suppose that a Killing vector field $K$ is timelike on a subset $\Wc\subset\Mc$. If there exists a three-dimensional hypersurface such that each integral curve of $K$ passing through $\Wc$, intersects {\em orthogonally} the hypersurface, then $\Wc$ is called {\em static}\footnote{If an open set $\Wc$ is static, then on a neighborhood of each point in $\Wc$ there exists a coordinate system $(x^\alpha)=(x^0,x^i)_{i=1,2,3}$ such that $K=\partial_{x^0}$, all the components $g_{\alpha\beta}$ of the metric in this system do not depend on $x^0$, and all the components $g_{0i}$ are zero.} \cite{wald}. An integral curve of such a Killing vector field, is the world line of a {\em static} observer. 
  
We will say that some (at least two) observers are  {\em Killing observers} if their world lines are integral curves of  {\em the same} Killing vector field, defined on a connected set $\Uc$. Note that this definition does not require the field $K$ to be timelike on the whole $\Uc$. Obviously, each Killing observer is stationary. 

We will present a physical interpretation of Killing observers in Section \ref{rigid}.

\subsubsection{Derivation of the time dilation ratio \label{der}}

Consider a Killing vector field $K$, defined on a subset of a spacetime $\Mc$, and a pair of Killing observers $\Oc$ and $\Oc'$ given by $K$. Let $\{\phi_s\}$ be the flow of $K$, and let 
\begin{align*}
\lambda&\mapsto \kappa(\lambda), & \lambda&\mapsto \kappa'(\lambda)
\end{align*}     
be integral curves of $K$, which coincide with world lines of the observers $\Oc$ and $\Oc'$, respectively. Let us now calculate the gravitational time dilation ratio between the observers.  

\paragraph{Choice of intervals}  Let us choose arbitrarily two consecutive events $\z_1$ and $\z_2$, which lie on $\kappa$, and which define an interval $\kappa_{\z_1\z_2}\subset\kappa$. Then there exists a number $s_0$ such that
\begin{equation}
\z_2=\phi_{s_0}(\z_1).
\label{zz-Ds}
\end{equation}

Suppose now that there exists a geodesic $\gamma_1$, which begins at $\z_1$, and ends at an event $\z'_1$ on $\kappa'$. Since $\phi_{s_0}$ preserves the metric $g$ (see Equation \eqref{phisg-g}), the curve
\begin{equation}
\gamma_2=\phi_{s_0}(\gamma_1)
\label{gg-Ds}
\end{equation}
is also a geodesic. Clearly, $\gamma_2$ begins at $\z_2$ and ends at an event
\begin{equation}
\z'_2=\phi_{s_0}(\z'_1),
\label{z'z'-Ds}
\end{equation}
which belongs to $\kappa'$.
 
In this way we obtained two events $\z'_1$ and $\z'_2$, which define an interval $\kappa'_{\z'_1\z'_2}\subset\kappa'$. Now it is enough to calculate the lapses $\Delta\tau$ and $\Delta\tau'$ of time along, respectively, $\kappa_{\z_1\z_1}$ and  $\kappa'_{\z'_1\z'_2}$, and the ratio \eqref{dil}.

\paragraph{Calculations} Suppose that Equations \eqref{z-lambda} hold. Then by virtue of \eqref{D-tau} 
\begin{equation}
\Delta \tau=\int^{\lambda_2}_{\lambda_1}\sqrt{|g(\dot{\kappa},\dot{\kappa})|}\,d\lambda.
\label{D-tau-int}
\end{equation}
But $\lambda\mapsto\kappa(\lambda)$ is an integral curve of $K$, and therefore the tangent vector 
\[
\dot{\kappa}(\lambda)=K(\kappa(\lambda)),
\]
and consequently
\[
g(\dot{\kappa},\dot{\kappa})=g(K,K).
\]

An important observation is that $g(K,K)$ is constant along each integral curve of $K$. To prove this, note that by shifting points along an integral curve of $K$, we do not change the curve\footnote{Strictly speaking, the action of the map $\phi_s$ amounts to a simple reparameterization of the curve, but the reparameterized curve remains an integral curve of $K$.}. Consequently, every map $\phi_s$ in the flow of $K$ preserves all integral curves of $K$, and therefore $K$ is preserved by every $\phi_s$: 
\begin{equation}
\phi_{s*}K=K.
\label{phiK-K}
\end{equation}
This fact together with \eqref{phisg-g} imply that both $K$ and $g$ do not change, when shifted along the integral curves of $K$. This is why $g(K,K)$ is constant along the curves.

The result just proven means that $g(\dot{\kappa},\dot{\kappa})$ in \eqref{D-tau-int} does not depend on the parameter $\lambda$. Denoting
\[
g(\dot{\kappa},\dot{\kappa})\equiv g_{KK},
\]
we obtain  
\[
\Delta \tau=\int^{\lambda_2}_{\lambda_1}\sqrt{|g_{KK}|}\,d\lambda= \sqrt{|g_{KK}|}\,(\lambda_2-\lambda_1)=\sqrt{|g_{KK}|}\,s_0,
\]
where the last equality follows from \eqref{z-lambda} and \eqref{zz-Ds}.

Similarly, $g(\dot{\kappa}',\dot{\kappa}')\equiv g'_{KK}$ is constant along $\kappa'$, and therefore
\[
\Delta\tau'=\sqrt{|g'_{KK}|}\,s_0.
\] 

Thus we arrive at the well-known formula, describing the gravitational time dilation ratio between a pair of Killing observers: 
\begin{equation}
D=\frac{\Delta\tau'}{\Delta\tau}=\sqrt{\frac{|g'_{KK}|}{|g_{KK}|}}.
\label{D-stat}
\end{equation}

\subsubsection{Discussion \label{disc}}

\paragraph{Distinguished features of time dilation between Killing observers} Since both $g_{KK}$ and $g'_{KK}$ are constant along the world lines of the stationary observers $\Oc$ and $\Oc'$, the time dilation ratio \eqref{D-stat} does not depend on the choice of events $\z_1$ and $\z_2$, which were the points of departure for the derivation of the ratio. 

The ratio \eqref{D-stat} is also independent of the choice of the geodesic $\gamma_1$. When deriving the ratio, we did not assume that the geodesic is null, which means that it can be timelike, null or spatial. And since $\phi_{s_0}$ preserves the metric $g$, the geodesic $\gamma_2$ given by \eqref{gg-Ds}, belongs to the same class as $\gamma_1$.  Thus in order to associate an interval $\kappa'_{\z'_1\z'_2}$ with $\kappa_{\z_1\z_2}$, the observer $\Oc$ can send to the observer $\Oc'$ either a pair of free massive particles, or a pair of light flashes (or even a pair of free tachyons, if they only existed), and this choice does not affect the resulting ratio. It does not mean however that given the interval $\kappa_{\z_1\z_2}$, the resulting interval $\kappa'_{\z'_1\z'_2}$ is unique---there is still an ambiguity here as large as in the general case, but this ambiguity disappears, once we calculate the lapse $\Delta\tau'$ of time along each such $\kappa'_{\z'_1\z'_2}$. Therefore one can claim that in the case of the observers $\Oc$ and $\Oc'$,  there is {\em a natural correspondence} between the value $\Delta\tau$ and the value $\Delta\tau'$. 

It follows that, contrary to the general case, in order to establish the value of the time dilation ratio \eqref{D-stat}, it is enough to specify a pair of Killing observers.

Note also that the equality \eqref{gg-Ds} means that, while pairing the intervals $\kappa_{\z_1\z_2}$ and  $\kappa'_{\z'_1\z'_2}$, the second light flash (free massive particle) was sent by the observer $\Oc$ exactly in the same way, as the first one. This is because both the metric on a neighborhood of $\z_2$, and the initial four-momentum of the second light flash (particle), can be obtained by means of $\phi_{s_0}$ from the metric on a neighborhood of $\z_1$, and the initial four-momentum of the first light flash (particle)---in this sense, the relation between the initial four-momentum at $\z_2$, and the metric around $\z_2$, is the same as the relation between the initial four-momentum at $\z_1$, and the metric around $\z_1$. In particular, the equality \eqref{gg-Ds} does not allow the static observer $\Oc$ in the Schwarzschild spacetime, considered in Section \ref{rem}, to send the second light flash  along the non-radial geodesic if the first light flash was sent along the radial one\footnote{When deriving the time dilation ratio \eqref{D-stat}, we actually did not use the fact that $\gamma_1$ is a geodesic---to obtain the same ratio we can allow $\gamma_1$ to be any curve connecting the world lines of the observers, provided \eqref{gg-Ds} holds. Such a freedom would be, however, troublesome from the point of view of the experiment---if e.g., we would like $\gamma_1$ to be a non-geodesic timelike curve, then the observer $\Oc$ would have to sent to $\Oc'$ two non-free massive particles, and in order to satisfy the requirement \eqref{gg-Ds}, he would have to control the accelerations of both particles at all times during their travel. But if $\gamma_1$ is a null (timelike) geodesic, then $\Oc$ has to guarantee merely that the initial four-momenta of both light flashes (free massive particles), are appropriately correlated.}.     

\paragraph{Undefined time dilation} Let us remark that there exist pairs of Killing observers, where their world lines cannot be connected by any causal (timelike or null) curve. Even though in such a case we can still calculate the r.h.s. of \eqref{D-stat}, we should refrain from  interpreting it as the ratio of time dilation, since the observers can send to each other neither light flashes nor massive particles. 

A simple case of such a pair is provided by a Killing vector field in the Minkowski spacetime, generated by Lorentz boosts. If $(t,x,y,z)$ is a standard coordinate system on the spacetime, then boosts in the $x$ direction parameterized by the ``hyperbolic angle'' $\psi$, define a family of curves 
\[
\psi\mapsto \big(t(\psi),x(\psi),y(\psi),z(\psi)\big)=(t_0\cosh\psi+x_0\sinh\psi, t_0\sinh\psi+x_0\cosh\psi,y_0,z_0),
\]
where $t_0,x_0,y_0,z_0$ are constants. These curves are integral curves of a vector field
\begin{equation}
K=x\partial_t+t\partial_x,
\label{K-u-acc}
\end{equation}
and the flow of $K$ consists of the Lorentz boosts. Since the boosts preserve the Minkowski metric, $K$ is a Killing vector field. One can also show that timelike integral curves of $K$, are world lines of uniformly accelerated observers in the Minkowski spacetime.

The set, on which $K$ is timelike, is given by the inequality $x^2>t^2$, and is a disjoint union of two so called Rindler ``wedges''---the $x$ coordinate on one ``wedge'' is positive, and on the other one is negative. It is well known that there does not exist any causal curve, which begins in one ``wedge'', and ends in the other. Thus, if the world line of a uniformly accelerated observer $\Oc$ lies in one ``wedge'', and the world line of another such observer $\Oc'$ lies in the other, then the world lines cannot be linked by any causal curve. 

However, if a Killing vector field is timelike on a connected set $\Wc$, and if the world lines of two corresponding Killing observers pass through $\Wc$ sufficiently close to each other, then the world lines can be connected by both a timelike geodesic and a null one---this can be proved by means of Riemann normal coordinates\footnote{A definition of Riemann normal coordinates can be found e.g. in \cite{wald}.}.

In the rest of the paper, we will tacitly assume that Killing observers can send both light flashes and free massive particles to each other.

\paragraph{Gravitational redshift} If $\Oc$ and $\Oc'$  are Killing observers, then the time dilation ratio \eqref{D-stat} appears also in the gravitational redshift formula 
\begin{equation}
\frac{\nu}{\nu'}=\sqrt{\frac{|g'_{KK}|}{|g_{KK}|}},
\label{red-sh}
\end{equation}
which describes the relationship between the frequency $\nu$ of a photon sent by $\Oc$ towards $\Oc'$, and the frequency $\nu'$ of the same photon detected by $\Oc'$ (see e.g. \cite{wald}). This means that conclusions we are going to draw, concerning gravitational time dilation, can be easily translated into conclusions concerning gravitational redshift. 

\subsubsection{Conclusions}

 It follows from Equations \eqref{D-stat} and \eqref{red-sh}, that in the case of Killing observers given by a Killing vector field $K$, both phenomena: {\em gravitational time dilation and gravitational redshift appear if and only if $g_{KK}$ is not a constant function on the spacetime}. If in these equations, $|g'_{KK}|$ is greater than $|g_{KK}|$, then $\Delta\tau'$ is greater than $\Delta\tau$, and $\nu'$ is smaller than $\nu$. Therefore 
\begin{enumerate}
\item the time of the Killing observers always ``slows down'' as $|g_{KK}|$ decreases;
\item from the point of view of the Killing observers, a photon always gets redshifted if $|g_{KK}|$ increases along its world line. 
\end{enumerate}

\subsection{Two measures of strength of gravity \label{two-m}}

The title question can be now formulated as follows: Is there a correlation between $|g_{KK}|$ and strength of the gravitational field, which would justify the statement that ``time always slows down as gravity increases''? More precisely: does $|g_{KK}|$ always decrease together with increases in the value of a scalar field, chosen to measure strength of gravity, assuming the Einstein equations (with a ``physically reasonable'' energy-momentum tensor) are satisfied? 

Of course, the answer to this question would be in affirmative if one treated $|g_{KK}|^{-1}$ as a measure of strength of gravity, as it is done in the textbook \cite{anderson}, in a discussion of the gravitational redshift formula \eqref{red-sh}. However, $|g_{KK}|^{-1}$ being the reciprocal of one of several independent components of the metric, is a rather poor measure of strength of the gravitational field (unless the field is approximately Newtonian \cite{wald}). Therefore, for this purpose we will use other scalar fields derived from the metric. 

One of these scalar fields, is the square of the Riemann tensor called the {\em Kretschmann scalar} 
\[
R^\alpha{}_{\beta\mu\nu}R_{\alpha}{}^{\beta\mu\nu}\equiv \Rc^2.
\]
In many spacetimes, this scalar field is positive everywhere and blows up to infinity, while approaching a singularity. Therefore it is quite natural to use it as a measure of strength of gravity\footnote{The Kretschmann scalar is not a flawless measure of strength of gravity, since there exists a solution of the Einstein vacuum equations of non-zero curvature, for which the scalar is zero \cite{kret-zero}---thus the strength of gravity as measured by the Kretschmann scalar, is the same in a spacetime described by this solution, and in the Minkowski spacetime.}.  

Another scalar field, which can serve as such a measure, is the square of the four-acce\-le\-ra\-tion of stationary observers. Let $K$ be a timelike Killing vector field defined on a subset $\Wc\subset\Mc$. Then each integral curve of $K$ can be treated as the world line of a stationary observer. The four-velocity $U$ of such an observer reads as
\begin{equation}
U=\frac{1}{\sqrt{|g_{KK}|}}K.
\label{U}
\end{equation}
The four-acceleration $A$ of the observer, is the covariant derivative of $U$ along itself,
\[
A=\nabla_UU,
\]
and can be thought of as an acceleration (or a force per unit mass) of a non-gravitational origin, needed to keep the observer on his stationary world line. On the other hand, by virtue of the equivalence principle, the gravitational acceleration experienced by free bodies in the observer's (non-inertial) reference frame, is given by $-A$. Therefore the square
\[
g(A,A)\equiv g_{AA}
\]
can be treated as a measure of strength of gravity (as we will show below, $g_{AA}$ is positive if only $A\neq 0$). Let us emphasize that this measure depends on the choice of the timelike Killing vector field $K$,  which in many stationary spacetimes is not unique.

\subsection{Closer look at the four-acceleration \label{look}}

The four-acceleration $A$ of Killing observers will be the main tool, we will use below in order to answer the title question. Therefore it is necessary to take a closer look at it.

\paragraph{Explicit expression for $A$} Let us first find an explicit expression for $A$ in terms of components of the metric $g$. An elegant way to do this \cite{ludv}, is to use the following property of the Killing vector field $K$: components $K_{\alpha}$ of a one-form obtained from $K$ by lowering its index, satisfy (see e.g. \cite{wald,red-kill})
\begin{equation}
K_{\alpha;\beta}=-K_{\beta;\alpha},
\label{K-K}
\end{equation}
where the semicolon denotes the covariant derivative. Taking into account Equation \eqref{U}, we have   
\begin{multline*}
A^\mu=(\nabla_UU)^\mu=U^\mu{}_{;\beta}U^\beta=\Big(\frac{1}{\sqrt{|g_{KK}|}}K^\mu\Big){}_{;\beta}\frac{1}{\sqrt{|g_{KK}|}}K^\beta=\\=\Big(\frac{1}{\sqrt{|g_{KK}|}}\Big){}_{,\beta}K^\beta K^\mu\frac{1}{\sqrt{|g_{KK}|}}+\frac{1}{|g_{KK}|}K^\mu{}_{;\beta}K^\beta=\frac{1}{|g_{KK}|}K^\mu{}_{;\beta}K^\beta,
\end{multline*}
where the last equality holds by virtue of constancy of $g_{KK}$ along the integral curves of $K$. Now, by virtue of \eqref{K-K}, we have 
\begin{equation}
A^\mu=\frac{1}{|g_{KK}|}g^{\mu\alpha}K_{\alpha;\beta}K^\beta=-\frac{1}{|g_{KK}|}g^{\mu\alpha}K_{\beta;\alpha}K^\beta
\label{A-mu}.
\end{equation}
On the other hand,
\begin{multline*}
g_{KK,\alpha}=(g_{KK})_{;\alpha}=(g_{\gamma\beta}K^\gamma K^\beta )_{;\alpha}=g_{\gamma\beta;\alpha}K^\gamma K^\beta +g_{\gamma\beta}K^\gamma{}_{;\alpha} K^\beta+g_{\gamma\beta}K^\gamma K^\beta{}_{;\alpha}=\\=K_{\beta;\alpha} K^\beta+K^\gamma K_{\gamma;\alpha}=2K_{\beta;\alpha} K^\beta
\end{multline*}
($g_{\gamma\beta;\alpha}=0$ by definition of the covariant derivative). Inserting this result into \eqref{A-mu}, we thus obtain that
\begin{equation}
A=A^\mu\partial_{x^\mu}=-\frac{1}{2}\frac{g_{KK,\alpha}}{|g_{KK}|}g^{\mu\alpha}\partial_{x^\mu}=\frac{1}{2}\frac{g_{KK,\alpha}}{g_{KK}}g^{\alpha\mu}\partial_{x^\mu}.
\label{acc}
\end{equation}

The expression above allows us to find an explicit formula for $g_{AA}$: 
\begin{equation}
g_{AA}=g_{\alpha\beta}A^\alpha A^\beta=\frac{1}{4}\frac{g_{KK,\alpha}g_{KK,\beta}g^{\alpha\beta}}{g^2_{KK}}.
\label{gAA}
\end{equation}

\paragraph{$A$ and the flow of $K$} In Equation \eqref{acc} there appear the scalar field $g_{KK}$ and components $g^{\alpha\mu}$ of the ``inverse'' metric. Furthermore, $g_{KK}$ does not change, when shifted along integral curves of $K$. The ``inverse'' metric is completely determined by the metric $g$, and since the latter one is constant along the integral curves, the former one has to be constant too. We then see that $A$ is built from fields, which are preserved\footnote{Regarding the partial derivatives $g_{KK,\alpha}$ in Equation \eqref{acc}: the derivatives are  components of the one-form $dg_{KK}=g_{KK,\alpha}dx^\alpha$. A basic property of the exterior derivative $d$ is that it commutes with the action (pull-back) of every $\phi_s$. Therefore, if $\phi_s$ does not change $g_{KK}$, then it does not change its derivative $dg_{KK}$. Thus $A$ is built from $g_{KK}$, $dg_{KK}$ and the ``inverse'' metric, that is, from fields preserved by every $\phi_s$.} by all maps in the flow $\{\phi_s\}$ of $K$, and therefore all the maps preserve $A$, that is to say
\begin{equation}
\phi_{s*}A=A.
\label{phisA}
\end{equation}
This equation and Equation \eqref{phisg-g} together imply that the square $g_{AA}$ is constant along the integral curves of $K$, that is, along world lines of corresponding Killing observers.

\paragraph{$A$ is spacelike} If $A\neq 0$ at an event, then it is spacelike there. To justify this statement, let us recall that $U$ is normed: $g(U,U)=-1$. Calculating the covariant derivative of this equation with respect to $U$, we obtain  
\[
0=\nabla_U(g(U,U))=(\nabla_U g)(U,U)+g(\nabla_UU,U)+g(U,\nabla_UU)=2g(A,U),
\]
since the covariant derivative annihilates the metric: $\nabla_Ug=0$. Consequently, if $A\neq 0$, then $A$ must be spacelike,
\begin{equation}
g_{AA}>0,
\label{gAA>0}
\end{equation}
as a vector orthogonal to the timelike vector $U$. Consequently, $g_{AA}$ is in general a non-negative scalar field on $\Mc$.  

\subsection{Direction of growth of $|g_{KK}|$ \label{dir-gkk}}

As we already know, the time measured by Killing observers ``slows down'' together with decreases of the value of $|g_{KK}|$. It turns out that there is a simple criterion, which tells us, in which direction $|g_{KK}|$ increases or decreases. Let us calculate $A|g_{KK}|$, that is, the derivative of $|g_{KK}|$ in the direction of the four-acceleration $A$---using \eqref{acc} and \eqref{gAA} we obtain the following: 
\[
A|g_{KK}|=-g_{KK,\beta}A^\beta=-\frac{1}{2}\frac{g_{KK,\alpha}}{g_{KK}}g^{\alpha\beta}g_{KK,\beta}=-2g_{KK}\frac{1}{4}\frac{g_{KK,\alpha}g_{KK,\beta}g^{\alpha\beta}}{g^2_{KK}}=2|g_{KK}|g_{AA}.
\] 
We have just proven that if only $A$ is non-zero, then $g_{AA}$ is positive (see \eqref{gAA>0}). Thus for every non-zero $A$
\begin{equation}
A|g_{KK}|>0.
\label{Agkk}
\end{equation}

This result  means that $|g_{KK}|$ always increases in the direction of $A$,  and decreases in the opposite one---this statement is general, i.e., it is {\em valid for every timelike Killing vector field}, and does not require, e.g., that the metric admitting the vector field, satisfies the Einstein equations with a ``physically reasonable'' energy-momentum tensor. 

Consider now two observers $\Oc$ and $\Oc'$, defined by a Killing vector field $K$. Assume that their world lines, respectively, $\kappa$ and $\kappa'$, can be connected by an integral curve $\alpha$ of $A$. The inequality \eqref{Agkk} implies that if $A$ is non-zero on the interval of $\alpha$ between the world lines, and points out the direction from $\kappa$ to $\kappa'$, then 
\begin{enumerate}
\item the time of $\Oc$ ``runs slower'' than the time of $\Oc'$;
\item a photon sent by $\Oc$ to $\Oc'$ gets redshifted.  
\end{enumerate}

\subsection{Killing observers as a rigid body \label{rigid}}

At this moment we already have all the tools needed to attack the problem defined by the title question. However before that, we would like to present a physical interpretation of Killing observers, which thus far have appeared only as a geometrical notion. This interpretation will help us to understand one of the non-standard examples of time dilation described below. 

Let us consider a rigid body in Newtonian mechanics, and assume that it is rotating with a constant angular velocity $\vec{\omega}$ with respect to an inertial frame. In a non-inertial  reference frame fixed to the body
\begin{enumerate}
\item points of the body do not change their positions;
\item points of the body are at rest with respect to each other---distances between all pairs of points do not change over time; 
\item the acceleration of each point of the body, is constant over time.
\end{enumerate} 

We will show below that for every timelike Killing vector field, there exists a local reference frame (understood as a coordinate system \cite{gr}), such that the corresponding Killing observers seen from this reference frame, possess properties analogous to the properties of the rigid body points listed above.

\begin{figure}
\psfrag{S}{$\Sigma_0$}
\psfrag{Sx}{$\Sigma_{s}$}
\psfrag{p}{$\phi_{s}$}
\psfrag{h}{$\kappa$}
\psfrag{x}{$x^i$}
\begin{center}
\includegraphics{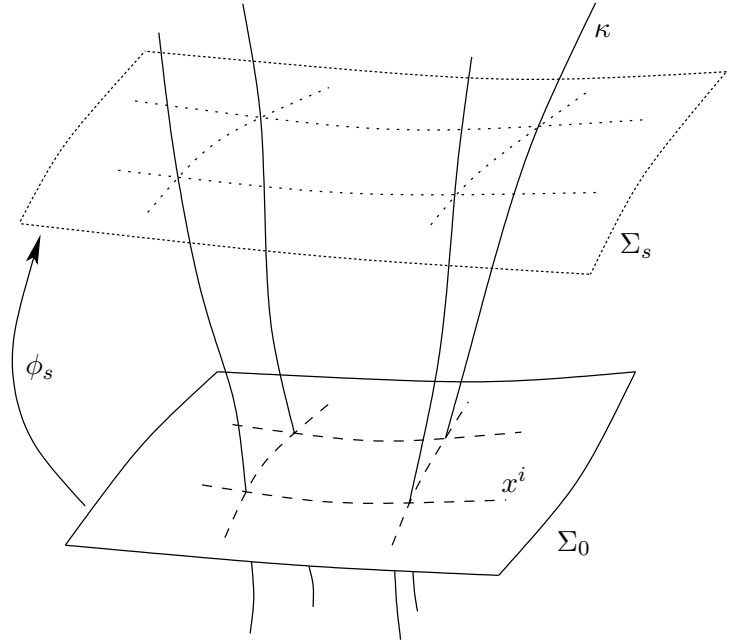}
\end{center}
\caption{Construction of the coordinate system $(x^0,x^i)$ \label{coor}}
\end{figure}

Consider a timelike Killing vector field $K$, and a {\em spacelike} three-dimensional hypersurface $\Sigma_0$ such that each point of $\Sigma_0$ belongs to an integral curve\footnote{We, however, do not assume that each integral curve of $K$ intersects $\Sigma_0$---our construction is local.} of $K$---see Figure \ref{coor}, where we showed four integral curves of the vector field, and denoted one of them by $\kappa$. Assume that $\Sigma_0$ is covered by a coordinate system $(x^i)_{i=1,2,3}$, depicted in Figure \ref{coor} by dashed lines. Using an element $\phi_{s}$ of the flow of $K$, we map $\Sigma_0$ to a hypersurface $\Sigma_{s}$ together with the coordinate system $(x^i)$. Using in this way all elements of the flow, we obtain a local coordinate system near $\Sigma_0$---a point $\z$ has coordinates $(x^\alpha)\equiv(x^0,x^i)$ if it belongs to $\Sigma_{x^0}$, and if its coordinates on the hypersurface are $(x^i)$.

Note now that the coordinates $(x^i)$ are constant along each integral curve $\kappa$ of $K$ intersecting $\Sigma_0$, and that $x^0$ is a parameter along $\kappa$. Therefore in the coordinate system $K=(1,0,0,0)=\partial_{x^0}$, and by virtue of \eqref{phiK-K}, the vector field $\partial_{x^0}$ is preserved by the flow $\{\phi_s\}$ of $K$. Since the remaining coordinates $(x^i)$ on the vicinity of $\Sigma_0$ were obtained by means of the flow, each vector field $\partial_{x^i}$ is preserved by the flow as well. Thus for every $\phi_s$      
\[
\phi_{s*}\partial_{x^\alpha}=\partial_{x^\alpha}.
\]

This fact together with \eqref{phisg-g} and \eqref{phisA}, mean that in the coordinate system $(x^\alpha)$,  {\em the components of the metric $g$ and the components of the four-acceleration $A$,  are constant} along the integral curves of $K$, that is, they do not depend on the coordinate $x^0$.          

The metric $g$ in the system $(x^\alpha)$ reads then as follows:
\begin{equation}
g=g_{KK}(x^k)(dx^0)^2+2g_{0i}(x^k)dx^0dx^i+g_{ij}(x^k)dx^idx^j.
\label{g-kill}
\end{equation}
This expression implies that $x^0$ is a timelike coordinate.  Since $\Sigma_0$ was chosen to be spatial, the coordinates $(x^i)$ on it are also spatial, and because $g_{ij}$ above does not depend on $x^0$, the coordinates $(x^i)$ are spatial everywhere. 

We already know that the coordinates $(x^i)$ are constant along the world lines of Killing observers defined by $K$. Thus from the point of view of the coordinate system/reference frame $(x^\alpha)$, each observer stays always at a position $(x^i)$. Because the coordinates $(x^i)$ are spatial, in any instant $x^0$, the distances between the observers in the reference frame, are defined by the last term in \eqref{g-kill}. Since this term is constant over time $x^0$, the distances do not depend on the time either---the observers are at rest with respect to each other. Moreover, the components of $A$ in the system $(x^\alpha)$ do not depend on $x^0$, and therefore the four-acceleration of each Killing observer perceived from the system, does not change over the time.

We thus see that Killing observers do indeed share some essential features with the points of a rigid body, rotating with a constant angular velocity. Therefore we can think of Killing observers as a GR generalization of a Newtonian rigid body.
 
\section{Gravitational time dilation and strength of gravity \label{td-str}}

To answer the title question, we will consider a number of spacetimes admitting a (local) timelike Killing vector field. The metric on each spacetime will satisfy either the vacuum Einstein equations (possibly with a cosmological constant), or the Einstein--Maxwell equations. Given such a spacetime with a fixed timelike Killing vector field, we will find the four-acceleration $A$ of the corresponding stationary observers, its square $g_{AA}$, and the Kretschmann scalar $\Rc^2$. We will then check, how $g_{AA}$ and $\Rc^2$ change along the vector field $A$ by calculating the derivatives of the scalar fields in the direction of $A$. 

Suppose that in some case it turns out that the derivative $A\Rc^2$ is negative. This together with \eqref{Agkk}, will mean that $\Rc^2$ decreases in the direction of growth of $|g_{KK}|$, or equivalently, that $\Rc^2$ increases in the direction of decrease of $|g_{KK}|$. Since the time of Killing observers ``slows down as $|g_{KK}|$ decreases'', we will conclude that in the case under consideration, the time ``slows down as gravity (measured by $\Rc^2$) increases''. A similar conclusion will be drawn if it will turn out that $Ag_{AA}<0$.

On the other hand, each case where 
\begin{align}
A\Rc^2&\geq0 \quad \quad \quad \text{or} \quad \quad \quad Ag_{AA}\geq0
\label{AR>0}
\end{align}
will be treated as an argument against the claim that ``time always slows down as gravity increases''.

As mentioned in the introduction, it is not difficult to find examples of Killing observers, for which at least one of the inequalities \eqref{AR>0} is satisfied. To keep the size of the paper below a reasonable limit, we will present merely examples of this kind, which will teach us something important. We encourage the readers, who will feel disappointed by a small number of the examples, to do their own search (many stationary and static solutions of the Einstein equations can be found in \cite{exact-sol,exact-st}). 

\subsection{The Schwarzschild spacetime \label{sec-schw}}

As a warming-up exercise, and as a test of the tools constructed above, let us consider the Schwarzschild spacetime. In this setting we have: \emi the components of the metric \eqref{schw} do not depend on the coordinate $t$, and therefore $K=\partial_t$ is a Killing vector field, \emii for $r>2M$
\[
g_{KK}=-\Big(1-\frac{2M}{r}\Big)
\]
is negative, and hence $K$ is timelike there; \emiii if $r\neq 2M$, then $K$ is everywhere orthogonal to the three-dimensional hypersurfaces given by $t={\rm constant}$, since all the components $g_{tr}, g_{t\theta}, g_{t\varphi}$ are zero. Thus the region of the spacetime given by $r>2M$, is {\em static}. This region and this Killing vector field   provide a case of gravitational time dilation, considered very often in textbooks on GR. Obviously in this case ``time slows down as gravity increases''. 

The four-acceleration $A$ of static observers defined by $K$, points out the direction outward from the source of the gravitational field (being a physical singularity at $r=0$):
\begin{equation}
A=\frac{M}{r^2}\partial_r.
\label{A-schw}
\end{equation}
The two scalar fields chosen to measure strength of gravity, in this case read as follows: 
\begin{align}
{\cal R}^2&=\frac{48M^2}{r^6}, & g_{AA}&=\frac{M^2}{r^3(r-2M)}
\label{scal-schw}
\end{align}
and
\begin{align*}
A{\cal R}^2&=-\frac{288M^3}{r^9}<0, & Ag_{AA}&=-\frac{2M^3(2r-3M)}{r^6(r-2M)^2}<0,
\end{align*}
as expected.

\subsection{The Schwarzschild spacetime with the source of negative mass}

Note now that if we allow $M<0$ in the Schwarzschild metric, then the resulting spacetime is static everywhere, and the four-acceleration \eqref{A-schw} changes its direction to the opposite one. Furthermore, the expressions for the scalar fields \eqref{scal-schw} remain unchanged, and consequently  
\begin{align*}
A{\cal R}^2&=-\frac{288M^3}{r^9}>0, & Ag_{AA}&=-\frac{2M^3(2r-3M)}{r^6(r-2M)^2}>0.
\end{align*}
Thus in this case ``time slows down as gravity {\em decreases}''. 

Of course, the Schwarzschild metric with a negative mass, is unphysical. Considering such a case can seem to be a trick, but nevertheless, this case gives us an important lesson: this metric is a perfectly valid solution of the vacuum Einstein equations, which means that these equations by themselves do not guarantee that ``time slows down as gravity increases''.  

\subsection{Four-acceleration directed inward to a source of a positive mass}

Cases like that presented in the previous section, where a negative mass of a source of the gravitational field, causes the four-acceleration of stationary observers to point out the direction inward to the source, can be easily ruled out as unphysical. Here we will present two physically valid cases, where the four-acceleration of Killing observers points out the direction inward to a source of a positive mass.

\subsubsection{Orbiting stationary observers in the Schwarzschild spacetime \label{sec-schw-orb}}

\paragraph{Definition and analysis} Consider again the region $r>2M$ of the Schwarzschild spacetime with a positive mass $M$, and note that $\partial_\varphi$ is a spacelike Killing vector field of the metric \eqref{schw}. Properties of the Lie derivative appearing in \eqref{LK-g}, imply that a linear combination of Killing vector fields, is also such a vector field. Thus for any real number $\omega$, the vector field
\[
\bar{K}=\partial_t+\omega\partial_\varphi
\]
is a Killing vector field of the metric. Indeed, if 
\begin{align*}
\bar{t}&=t,& \bar{\varphi}&=\varphi-\omega t
\end{align*} 
then $K=\partial_{\bar{t}}$ in the coordinate system $(\bar{t},r,\theta,\bar{\varphi})$, and the metric \eqref{schw} can be rewritten in the following form
\begin{equation}
g=-\Big(1-\frac{2M}{r}-\omega^2r^2\sin^2\theta\Big)d\bar{t}^2+2\omega r^2\sin^2\theta\, d\bar{t}\,d\bar{\varphi}+\Big(1-\frac{2M}{r}\Big)^{-1}dr^2+r^2(d\theta^2+\sin^2\theta\,d\bar{\varphi}^2),
\label{schw-orb}
\end{equation} 
where the components do not depend on $\bar{t}$. 

Hence  $\bar{K}$  is timelike if 
\begin{equation}
g_{\bar{K}\bar{K}}=-1+\frac{2M}{r}+\omega^2r^2\sin^2\theta<0.
\label{g-bkbk}
\end{equation}
The formula \eqref{acc} applied to $g_{\bar{K}\bar{K}}$, gives 
\[
\bar{A}=\frac{1}{g_{\bar{K}\bar{K}}}\Big[\Big(1-\frac{2M}{r}\Big)\Big(\omega^2r\sin^2\theta-\frac{M}{r^2}\Big)\partial_r+\omega^2\sin\theta\cos\theta\,\partial_\theta\Big]
\]
provided $g_{\bar{K}\bar{K}}\neq 0$.

Our goal now is twofold: \emi to check, whether inequality \eqref{g-bkbk} can be satisfied for $r>2M$, and \emii to find the direction of $\bar{A}$ in the region, where \eqref{g-bkbk} is satisfied, that is, in the region, where $\bar{A}$ is the four-acceleration of stationary observers, defined by $\bar{K}$. To simplify the task we will restrict ourselves to the ``equatorial plane'' $\theta=\pi/2$, where
\begin{align}
g_{\bar{K}\bar{K}}&=-1+\frac{2M}{r}+\omega^2r^2,\label{g-bkbk-eq}\\
\bar{A}&=\frac{1}{g_{\bar{K}\bar{K}}}\Big(1-\frac{2M}{r}\Big)\Big(\omega^2r-\frac{M}{r^2}\Big)\partial_r\equiv \bar{A}^r\partial_r.\label{A-bar}  
\end{align}

Let us first consider the inequality \eqref{g-bkbk} on the ``equatorial plane'':
\begin{equation}
0>g_{\bar{K}\bar{K}}=-1+\frac{2M}{r}+\omega^2r^2.
\label{gkk-om}
\end{equation}
For $r>2M$ it is equivalent to
\[
0>r^3-\frac{r}{\omega^2}+\frac{2M}{\omega^2}\equiv W(r).
\]
To analyze this condition, let us treat $W$ as a polynomial, defined on the whole $\R$. Clearly, $\lim_{r\to-\infty}W(r)=-\infty$, and $W(2M)=8M^3>0$ independently of $\omega$. Hence there always exists a root $r_1<2M$ of the polynomial. Thus the polynomial cannot have negative values for any $r>2M$ if it does not have three pairwise distinct real roots.

Note that $W$ is of the canonical form
\[
W(r)=r^3+pr+q
\]
with $p=-\omega^{-2}$ and $q=2M\omega^{-2}$. There exist three pairwise distinct real roots of $W$ if
\[
0>q^2+\frac{4p^3}{27}=\frac{4M^2}{\omega^4}-\frac{4}{27\omega^6},
\]
which gives the following condition imposed on $\omega$: 
\begin{equation}
\omega^2<\frac{1}{27M^2}.
\label{omega2<}
\end{equation}

Now note that if
\[
r=\sqrt[3]{\frac{M}{\omega^2}}\equiv r_0,
\]
then $\bar{A}$ given by \eqref{A-bar} vanishes, because the factor $(\omega^2r-M/r^2)$ is zero at $r=r_0$. If the inequality \eqref{omega2<} holds, then
\[
W(r_0)=\frac{3M}{\omega^2}\Big(1-\frac{1}{\sqrt[3]{27M^2\omega^2}}\Big)<0, 
\]
and consequently there exists two other real roots $r_2,r_3$ of $W$  such that
\begin{equation}
2M<r_2<r_0<r_3. 
\label{rrr}
\end{equation}

Thus if the condition \eqref{omega2<} is satisfied, then $W(r)<0$ for every $r\in\,]r_2,r_3[$. Consequently, on the ``equatorial plane'' for these values of the radial coordinate $r$, the Killing vector field $\bar{K}$ is timelike, and $\bar{A}$ given by \eqref{A-bar} is the four-acceleration of corresponding stationary observers. It is easy to see that for any fixed $t_0$ and $\varphi_0$, and for any fixed $r\in\,]r_2,r_3[$, the curve  
\begin{equation}
\lambda\mapsto \big(t(\lambda),r(\lambda),\theta(\lambda),\varphi(\lambda)\big)=(\lambda+t_0,r,\pi/2,\omega \lambda+\varphi_0),
\label{wl}
\end{equation}
is an integral curve of $\bar{K}$, which describes the world line of a stationary observer---note that the radial coordinate $r$ of such an observer, is constant along his world line.

Moreover, the properties of the factor $(\omega^2r-M/r^2)$ in \eqref{A-bar}, imply that 
\begin{equation}
\bar{A}^r(r)
\begin{cases}
>0 & \text{if $r\in\,]r_2,r_0[$,}\\
=0 & \text{if $r=r_0$,}\\
<0 & \text{if $r\in\,]r_0,r_3[$}
\end{cases},
\label{Ar-0}
\end{equation}
where $\bar{A}^r(r)$ is the component of the four-acceleration \eqref{A-bar}. This means that: 
\begin{enumerate}
\item the four-acceleration of the observers of $r\in\,]r_2,r_0[$, points out the direction outward from the source of the gravitational field,
\item the observers of $r=r_0$ are free observers (their world lines are geodesics),
\item the four-acceleration of the observers of $r\in\,]r_0,r_3[$, points out the direction {\em inward to the source}.     
\end{enumerate}

\paragraph{Discussion of the results} To understand these results, recall first that they are valid on the ``equatorial plane'' only. Then note that for $r$ close to $2M$, the component $g_{tt}$ of the metric \eqref{schw} is close to zero and $g_{\varphi\varphi}>4M^2$, therefore the Killing vector field $\bar{K}=\partial_t+\omega\partial_{\varphi}$ must be spatial there. On the other hand, for large $r$, $g_{tt}$ is close to $-1$ and $g_{\varphi\varphi}\gg 1$, and again $\bar{K}$ has to be spatial there. This is why $\bar{K}$ cannot be timelike for all $r>2M$.      

It follows from \eqref{wl}, that each stationary observer under consideration,  moves with respect to the ``static'' coordinate system/reference frame $(t,r,\theta,\varphi)$ (and with respect to the source), along a circular orbit of ``radius'' $r$, with the constant angular velocity given by
\[
\frac{d\varphi}{dt}=\frac{d\varphi/d\lambda}{dt/d\lambda}=\omega.
\] 

In order to stay on their world lines, the stationary observers need to be subjected to the four-acceleration $\bar{A}$ given by \eqref{A-bar}. Furthermore, $\bar{A}$ can be rewritten as a sum 
\[
\bar{A}=\bar{A}_g+\bar{A}_c,
\]
where 
\begin{equation}
\begin{aligned}
\bar{A}_g&=|g_{\bar{K}\bar{K}}|^{-1}\Big(1-\frac{2M}{r}\Big)\frac{M}{r^2}\partial_r,\\
\bar{A}_c&=-|g_{\bar{K}\bar{K}}|^{-1}\Big(1-\frac{2M}{r}\Big)\omega^2r\partial_r.
\end{aligned}
\label{Ag-Ac}
\end{equation}
In both formulas above the factor $|g_{\bar{K}\bar{K}}|^{-1}$ comes from the normalization of the four-velocity $U$ (see Equation \eqref{U} and the derivation of the general expression for the four-acceleration in Section \ref{two-m}). The factor $(1-2M/r)$ in \eqref{Ag-Ac}, can be understood as a GR correction to the factors $M/r^2$ and $\omega^2r$, which describe respectively, the values of Newtonian gravitational acceleration, and Newtonian centripetal acceleration. The term $\bar{A}_g$ is directed outward from the source, and can be interpreted as the acceleration, needed to counteract the gravitational acceleration of the observer, originating from the source. On the other hand, the term $\bar{A}_c$ is directed inward to the source, and can be thought of as a centripetal acceleration, which causes the observer to move along his circular orbit.

In the case of the stationary observers in the orbit of $r=r_0$, the gravitational acceleration is equal exactly to the centripetal acceleration, required to keep them in the orbit, and therefore $\bar{A}$ is equal zero there. In orbits placed closer to the source, that is $r\in\,]r_2,r_0[$,  the gravitational acceleration is stronger than the centripetal one: $M/r^2>\omega^2r$. As a result, the value of $\bar{A}_g$ exceeds the value of $\bar{A}_c$, and $\bar{A}$ points out the direction outward from the source. On the other hand, in orbits placed farther from the source, that is $r\in\,]r_0,r_3[$, the situation is opposite: here the required centripetal acceleration is stronger than the gravitational one, and consequently the four-acceleration $\bar{A}$ of the corresponding stationary observers, points out the direction inward to the source.        
  
\paragraph{Conclusions} We have shown that the stationary observers given by the Killing vector field $\bar{K}$, which revolve around the source (of a positive mass) in orbits of $r\in\,]r_0,r_3[$ contained in the ``equatorial plane'', are distinguished by their four-acceleration $\bar{A}$, which is directed inward to the source. Therefore, the time of these observers ``slows down'' in orbits located {\em farther from the source}. Moreover, from the point of view of these observers, a photon moving {\em away from the source, gets blueshifted}.

It is clear that for the observers under consideration we have
\[
\bar{A}\Rc^2>0,
\]    
where $\Rc^2$ is the Kretschmann scalar in the Schwarzschild spacetime, given by the first equation \eqref{scal-schw}.

Unfortunately, the square $g_{\bar{A}\bar{A}}(r)$ (i.e. the square $g_{\bar{A}\bar{A}}$ restricted to the ``equatorial plane'') is a complicated rational function of $r$, and the required analysis of the sign of $\bar{A}g_{\bar{A}\bar{A}}(r)$, exceeds our calculating capacity. It is, however, easy to see that $\bar{A}g_{\bar{A}\bar{A}}(r)<0$ in the vicinity of $r_0$---this is because $g_{\bar{A}\bar{A}}(r)$ has a minimum at $r_0$, and the four-acceleration $\bar{A}$ on the ``equatorial plane'', always points out {the orbit of $r=r_0$} (see \eqref{Ar-0}).          

Note that on the ``equatorial plane'' 
\[
\lim_{r\to r^-_3} |g_{\bar{K}\bar{K}}|=0.
\] 
This fact, together with Equations \eqref{D-stat} and \eqref{red-sh}, mean that on the ``plane'', for every orbiting stationary observer $\Oc$, that is, for every observer in an orbit of $r\in\,]r_2,r_3[$, there exists another stationary observer $\Oc'$ {such that:
\begin{enumerate}
\item {$\Oc'$ orbits {\em farther} from the source than $\Oc$,}
\item {the time of $\Oc'$ ``runs {\em slower}'' than the time of $\Oc$,}
\item {a photon sent by $\Oc$ to $\Oc'$ gets {\em blueshifted}.}
\end{enumerate}}

\paragraph{Possible objection} Let us note that one could object to all these conclusions above, saying that time dilation perceived by the orbiting stationary observers is not solely gravitatio\-nal---these observers are not static, that is, they are not at rest with respect to the source, and the ``static'' coordinate system/reference frame $(t,r,\theta,\varphi)$ on the Schwarzschild spacetime. Therefore the time dilation ratio 
\begin{equation}
D=\sqrt{\frac{1-{2M}/{\r'}-\omega^2\r^{\prime 2}}{1-2M/\r-\omega^2\r^2}}
\label{D-orb}
\end{equation}
between observers in orbits of ``radius'' $r=\r'$ and $r=\r$, contained in the ``equatorial plane'', is {a cumulative result of two distinct phenomena}. {One of these phenomena, is} time dilation due to the relative velocity of the observers (related to the Doppler effect), and the other is ``true'' gravitational time dilation caused solely by the static gravitational field (such distinctions are made for instance in descriptions of relativity effects in the GPS, see e.g. \cite{gps}).  

Our reply to this objection will consist of three counterarguments.

Firstly, static gravitational fields are quite exceptional. Therefore the restriction of the term ``gravitational'' to static aspects of gravity, would mean that a lot of phenomena predicted by GR, including gravitational waves, are not gravitational. Moreover, in the case of stationary non-static spacetimes like the Kerr spacetime \cite{kerr}, there is no way to make any distinction between time dilation due to the relative velocity of stationary observers, and ``true'' gravitational time dilation. 

{Secondly, it is rather doubtful whether the contributions of the relative velocity and of the static gravitational field to the ratio \eqref{D-orb}, can be strictly defined or separated. To see this, let us consider the case $M=0$, that is, the Minkowski spacetime. In this case, the ``static'' reference frame $(t,r,\theta,\varphi)$ coincides with a global inertial frame, with spatial Cartesian coordinates replaced by the spherical ones. The static gravitational field, as seen from the frame, is trivial. Consequently, one may expect that in this case, there is no ``true'' gravitational time dilation between the orbiting stationary observers. Consider now two such observers, which move along the same orbit of radius $\r$, contained in the ``equatorial plane''. Suppose that at every instant, they are located on opposite sides of the orbit. Then in the inertial reference frame, the relative velocity of the observers is non-zero. Therefore one may expect that there is time dilation between the observers, due to the velocity. It seems then that  in the case of these two observers, the ratio \eqref{D-orb} should not be equal $1$ altogether. However, contrary to these expectations, the ratio is equal $1$.} 

Thirdly, from the point of view of an inertial frame in Newtonian mechanics, the relative velocities of many points of a rotating rigid body, are non-zero. But from the point of view of a reference frame fixed to the body, all the points are at rest with each other. Similarly, from the point of view of the ``static'' reference frame $(t,r,\theta,\varphi)$ (in the case $M\geq 0$), the relative velocities of the orbiting observers, seem to be non-zero. But the reference frame $(\bar{t},r,\theta,\bar{\varphi})$ (see \eqref{schw-orb}) is ``fixed to the rigid body'' of the orbiting observers, and from the point of view of this frame,  the observers are at rest with each other (see Section \ref{rigid}). Therefore the time dilation they perceive, is purely gravitational.

\subsubsection{The Schwarzschild--de Sitter spacetime}

The example of the orbiting observers, discussed above, indicates that as long as all {\em stationary} observers are considered, it is difficult to treat the statement ``time slow down as gravity increases'' as correct in general. But perhaps this statement would be correct without exceptions, if we restricted ourselves to {\em static} observers in {\em physically acceptable} spacetimes. Here we will show that this guess is not true.

A counterexample to the guess is provided by the Schwarzschild--de Sitter metric \cite{schw-desitt}:
\[
g=-\Big(1-\frac{2M}{r}-\frac{\Lambda}{3}r^2\Big)dt^2+\Big(1-\frac{2M}{r}-\frac{\Lambda}{3}r^2\Big)^{-1}dr^2+r^2(d\theta^2+\sin^2\theta\,d\varphi^2),\quad r>0,
\]  
where $M$ is the mass of the source of the gravitational field (being a physical singularity at $r=0$), and $\Lambda$ is a cosmological constant. Although this metric is a valid solution of the vacuum Einstein equations with the cosmological constant for any $M$ and $\Lambda$, we will consider only positive values of these constants. 

Obviously, $K=\partial_t$ is a Killing vector field. It is timelike if
\begin{equation}
g_{KK}=-1+\frac{2M}{r}+\frac{\Lambda}{3}r^2<0,
\label{gkk-la}
\end{equation}
and the region of the spacetime where this condition holds is static. 

Note now that since $\Lambda$ is assumed to be positive, $g_{KK}$  above is equal to $g_{\bar{K}\bar{K}}$ given by \eqref{g-bkbk-eq}, under the identification $\Lambda/3=\omega^2$. This means that the case under consideration, is similar to the formerly discussed case of the stationary observers, orbiting on the ``equatorial plane'' in the Schwarzschild spacetime. Therefore many (but not all) results concerning the present case, can be obtained by rewriting appropriately those derived in the previous section. 

Thus if 
\begin{align*}
&0<\Lambda<\frac{1}{9M^2}, & r_0&=\sqrt[3]{\frac{3M}{\Lambda}},
\end{align*}
then \eqref{gkk-la} holds on some interval $]r_2,r_3[$ (see also \cite{exact-st}), where $r_2,r_3$  satisfy \eqref{rrr}, and the four-acceleration of the static observers defined by $K$,    
\[
{A}=\Big(-\frac{\Lambda}{3} r+\frac{M}{r^2}\Big)\partial_r,
\]    
points out the direction inward to the source, provided $r\in\,]r_0,r_3[$.

Obviously, the static observers of $r\in\,]r_0,r_3[$, perceive time dilation and redshift exactly as the corresponding orbiting Killing observers, considered in the previous section.   

The Kretschmann scalar in the spacetime under consideration reads as
\[
\Rc^2=\frac{48M^2}{r^6}+\frac{8\Lambda^2}{3},
\]    
and if $r\in\,]r_0,r_3[$, then
\[
A\Rc^2>0.
\]
As in the previous case, the derivative  $Ag_{AA}$ is too complicated a function, to analyze its sign on the whole interval $]r_2,r_3[$, however, $Ag_{AA}<0$ on a vicinity of $r_0$.

Again, 
\[
\lim_{r\to r^-_3} |g_{{K}{K}}|=0.
\] 
This fact together with equations \eqref{D-stat} and \eqref{red-sh}, imply that for every static observer $\Oc$, that is, for an observer ``resting'' at $r\in\,]r_2,r_3[$, there exists another static observer $\Oc'$ {such that
\begin{enumerate}
\item {$\Oc'$ ``rests'' {\em farther} from the source than $\Oc$,}
\item {the time of $\Oc'$ ``runs {\em slower}'' than the time of $\Oc$,}
\item {a photon sent by $\Oc$ to $\Oc'$ gets {\em blueshifted}.}
\end{enumerate}}

\subsection{The Rindler spacetime \label{rind-st}}

In the Rindler spacetime \cite{rindler} the metric reads as
\begin{equation}
g=-\bar{x}^2d\bar{t}^2+d\bar{x}^2+dy^2+dz^2, \quad \bar{x}>0,
\label{rind}
\end{equation}
and $K=\partial_{\bar{t}}$ is a timelike Killing vector field such that
\begin{align*}
g_{KK}&=-\bar{x}^2, &  A&=\frac{1}{\bar{x}}\partial_{\bar{x}},\\
 g_{AA}&=\frac{1}{\bar{x}^2}, & Ag_{AA}&=-\frac{2}{\bar{x}^4}<0.
\end{align*}
This spacetime is flat and therefore
\[
A{\cal R}^2=0.
\]

Thus here the time ``slows down as gravity (measured by curvature) is {\em zero}'', and ``slows down as gravity (measured by $g_{AA}$) increases''.

In fact, the Rindler spacetime is a part of the Minkowski spacetime, and by a coordinate transformation \cite{rindler}, it can be identified with one of the Rindler ``wedges'', considered in Section \ref{disc}. The Killing vector field $K=\partial_{\bar{t}}$ coincides with the vector field \eqref{K-u-acc}, and therefore the stationary observers defined by $K=\partial_{\bar{t}}$, are uniformly accelerated observers in the Minkowski spacetime. 

\subsection{The Bertotti--Robinson spacetime}

So far we have encountered the inequality $Ag_{AA}\geq0$ only in the case of the Schwarzschild spacetime with a source of negative mass. This fact may suggest that perhaps $Ag_{AA}<0$ in every physically acceptable spacetime. Below we will present a counterexample to this conjecture. 

The Bertotti--Robinson metric \cite{ber,rob} is a solution of the Einstein--Maxwell equations, with a non-zero uniform electromagnetic field. Here we will use two distinct forms of the metric \cite{exact-st}:  
\begin{align*}
g&=\frac{e^2}{r^2}\big(-dt^2+dr^2+r^2(d\theta^2+\sin^2\theta\,d\varphi^2)\big)=\\
&=-\Big(1+\frac{\bar{r}^2}{e^2}\Big)d\bar{t}^2+\Big(1+\frac{\bar{r}^2}{e^2}\Big)^{-1}d\bar{r}^2+e^2(d\theta^2+\sin^2\theta\,d\varphi^2),
\end{align*}
where $e$ is a non-zero constant, and $r>0$. Clearly, the metric admits two timelike Killing vector fields $K=\partial_t$ and $\bar{K}=\partial_{\bar{t}}$. In the case of $K$, we obtain
\begin{align*}
g_{KK}&=-\frac{e^2}{r^2}, & A&=-\frac{r}{e^2}\partial_r,\\
g_{AA}&=\frac{1}{e^2}, & Ag_{AA}&=0,
\end{align*}     
and in the case of $\bar{K}$, we have
\begin{align*}
g_{\bar{K}\bar{K}}&=-\Big(1+\frac{\bar{r}^2}{e^2}\Big), & \bar{A}&=\frac{\bar{r}}{e^2}\partial_{\bar{r}},\\
g_{\bar{A}\bar{A}}&=\frac{\bar{r}^2}{e^2(e^2+\bar{r}^2)}, & \bar{A}g_{\bar{A}\bar{A}}&=\frac{2\bar{r}^2}{e^2(e^2+\bar{r}^2)^2}>0\quad \text{if $\bar{r}\neq 0$}.
\end{align*}
The Kretschmann scalar $\Rc^2$ in the Bertotti--Robinson spacetime, is constant. Thus
\begin{align*}
A\Rc^2&=0 \quad \quad \quad  {\rm and}  \quad \quad \quad   \bar{A}\Rc^2=0.
\end{align*}

We then see that in the case of static observers defined by $K$, the time ``slows down as gravity (measured by both $\Rc^2$ and $g_{AA}$) is {\em constant}''. The time of  static observers given by $\bar{K}$, ``slows down as gravity (measured by $\Rc^2$) is {\em constant}, and as gravity (measured by $g_{\bar{A}\bar{A}}$) {\em decreases}''.    

\subsection{Summary}

\begin{table}
\begin{center}
\begin{tabular}{||l|l|l|c|c||}
\hline \hline
Spacetime & Killing vector  & Killing & \multicolumn{2}{|c||}{Sign of}\\ \cline{4-5}
 \rule{0pt}{12pt} & field & observers & $A\Rc^2$ & $Ag_{AA}$\\
\hline
Schwarzschild, $M>0$ & $\partial_t$ & static & $-$ & $-$\\
\hline
 Schwarzschild, $M<0$ & $\partial_t$ & static & $+$ & $+$\\
\hline
Schwarzschild, $M>0$  & $\partial_t+\omega\partial_{\varphi}=\partial_{\bar{t}}$ & non-static & $\pm$ & ? \\    
\hline
Schwarzschild--de Sitter, $M,\Lambda>0$ & $\partial_t$ & static & $\pm$ & ? \\\hline
Minkowski (Rindler) & $x\partial_t+t\partial_x=\partial_{\bar{t}}$ & static & $0$ & $-$\\
\hline
Bertotti-Robinson & $\partial_t$ & static & $0$ & $0$ \\
\hline
Bertotti-Robinson & $\partial_{\bar{t}}$ & static & $0$ & $+$ \\
\hline\hline        
\end{tabular}
\end{center}
\caption{Analyzed examples of time dilation}
\label{tab}
\end{table}

The analyzed examples of gravitational time dilation between Killing observers, are summarized in Table \ref{tab}. Clearly, we encountered all possible types of behavior of $A\Rc^2$ and $Ag_{AA}$, in physically acceptable spacetimes (excluding the Schwarzschild one with $M<0$). We then conclude that the time dilation ratio \eqref{D-stat} is not correlated with the strength of gravity, measured by the scalar fields $\Rc^2$ and $g_{AA}$.  

\section{A no--go theorem}

In the Schwarzschild spacetime with $M>0$, the four-acceleration $A$ of the static observers, is {\em antiparallel} to the four-acceleration $\bar{A}$ of the orbiting stationary observers, of $r\in\,]r_0,r_3[$ and $\theta=\pi/2$ (see Sections \ref{sec-schw} and \ref{sec-schw-orb}). There is an important implication of this fact. 

To state this implication, recall first that in this paper, a measure of strength of gravity is understood as a scalar field, derived from the spacetime metric according to a prescription. Note now that we can distinguish two classes of such measures: observer independent measures and observer dependent ones. An example of an observer independent measure is the Kretschmann scalar $\Rc^2$---even if a metric $g$ admits (linearly) independent timelike Killing vector fields, then in order to calculate its  Kretschmann scalar $\Rc^2(g)$, we do not have to specify any of the vector fields, i.e., we do note have to specify any family of Killing observers. On the other hand, the measure $g_{AA}$ is obviously observer dependent.

{For the sake of a theorem, we are going to formulate below, let us be more precise. Namely, we define observer independent measure of strength of gravity, to be a map $F$ such that \emi its domain is the set of all Lorentzian metrics, defined on a four-dimensional open ball, and admitting timelike Killing vector fields,  and \emii its codomain is the set of all scalar fields on the ball. Then $F(g)$ is the value of the map $F$ on a metric $g$, i.e., $F(g)$ is a scalar field derived from $g$. Denote by $A$ the four-accele\-ra\-tion of observers, defined by a timelike Killing vector field $K$.}     
\begin{thr}
There is no observer independent measure $F$ of strength of gravity, such that for every metric $g$ admitting a timelike Killing vector field, and for every timelike Killing vector field $K$ of $g$,
\begin{equation*}
AF(g)<0
\end{equation*}
wherever $A\neq 0$. 
\label{thr-F}
\end{thr}
\noindent The theorem means that there does not exist an observer independent measure of the strength, which guarantees that ``time always slows down as gravity increases'' (see the beginning of Section \ref{td-str}).

The proof of the theorem is very simple: {if such a measure $F$ existed, then in the case of the Schwarzschild metric $g_{\rm \,Schw}$, the value of the scalar field $F(g_{\rm \,Schw})$, would have to decrease along both $A$ and $\bar{A}$, mentioned at the very beginning of the present section. But this is impossible, since the vector fields are antiparallel at some points}.  

\section{Redshift, time dilation and spacetime curvature}

In the paper \cite{miscon}, the authors cite some statements found in modern textbooks on GR, concerning a relation between gravitational redshift and spacetime curvature: according to these statements, the result of a {\em single} gravitational redshift experiment of local range, is a sufficient reason for introducing non-zero spacetime curvature. The authors rightly emphasize that such assertions are incorrect, arguing that in a laboratory accelerated with respect to inertial frames in the Minkowski spacetime, redshift is observed, but curvature is not involved. They claim however that
\begin{quote}
there {\em is} nonetheless, a connection between gravitational redshift results and spacetime curvature
\end{quote}
and make this connection precise, by saying that
\begin{quote}
{\em multiple} gravitational redshift experiments performed at different spatial locations (...), require for their joint explanation, the rejection of the global nature of inertial frames. 
\end{quote}

Unfortunately, the latter statement cannot be correct, since it is contradicted by, e.g., the uniformly accelerated observers in the Minkowski (Rindler) spacetime, considered in Section \ref{rind-st}: there measurements of non-trivial redshift, can be performed in many different spatial locations, but the spacetime is flat, and does admit global inertial frames. 

In fact, both gravitational redshift and gravitational time dilation, are rather common phenomena in the Minkowski spacetime, despite its zero curvature. Indeed, all Killing vector fields on this spacetime, form a ten-dimensional {vector space. In this space} there are many vector fields, which are locally timelike, and which define accelerated stationary observers---this non-zero four-acceleration does not allow the corresponding scalar field $g_{KK}$ to be constant (see Inequality \eqref{Agkk}), which guarantees the appearance of redshift and time dilation. One can then claim that  gravitational redshift and gravitational time dilation, are also {\em special relativity phenomena}.

Consider, on the other hand, the famous Einstein static universe \cite{ein-stat}:  
\begin{equation}
g=-dt^2+R^2\big(d\beta^2+\sin^2\beta(d\theta^2+\sin^2\theta\, d\varphi^2)\big),
\label{stat-Ein}
\end{equation}
where $R$ is a positive constant. Here for the timelike Killing vector field $K=\partial_t$, we have
\[
g_{KK}{=-1.}
\]
{Therefore} there is no redshift nor time dilation between static observers defined by $K$, despite the fact that the curvature of the metric \eqref{stat-Ein}, is non-zero \cite{exact-st}.

Thus spacetime curvature is not necessary for the appearance of gravitational redshift and gravitational time dilation, and does not force them to occur. 

There is a simple explanation of these {facts. We} know already that redshift and time dilation appear if and only if $g_{KK}$ is not constant, that is, if and only if at least one of the first partial derivatives of $g_{KK}$, does not vanish (except on lower dimensional hypersurfaces where $g_{KK}$ may take its locally minimum or maximum values). On the other hand, the Riemann curvature tensor is built from the first and second derivatives of all components of the metric. Then it is not a surprise that the non-zero first derivatives of $g_{KK}$, being just one component of the metric, do not guarantee that the curvature is non-zero. Moreover, non-zero curvature does not imply that the first derivatives of a particular component of the metric, do not vanish.    

To summarize: contrary to the claim of the authors of \cite{miscon}, there is no close connection between spacetime curvature, and both gravitational redshift and gravitational time dilation.

\section{``Accelerational'' time dilation}

Complete information about gravitational time dilation and gravitational redshift perceived by Killing observers, is contained in the scalar field $g_{KK}$. However, $g_{KK}$ may be too abstract a notion for some students and amateurs, and therefore teaching them these phenomena in terms of $g_{KK}$, may be not a good idea. From this point of view, the statement ``time slows down as gravity increases'' may be of some value, even if it is not correct in general. Nevertheless, we advocate to put this statement aside definitely, since basic facts about time dilation and redshift, can be strictly described in terms of the four-acceleration $A$ of the observers, which is a notion much more comprehensible than $g_{KK}$, and unambiguously defined, contrary to the notion of strength of gravity.

Indeed, as noted in the previous section, time dilation and redshift appear if and only if at least one of the first partial derivatives of $g_{KK}$, does not vanish. But by virtue of Equation \eqref{acc}, at least one of the derivatives does not vanish if and only if $A$ is non-zero. Thus both {\em gravitational time dilation and gravitational redshift, appear in the case of Killing observers if and only if their four-acceleration is non-zero}. In other words, these phenomena are conditioned by the four-acceleration, and taking into account a non-gravitational origin of the four-acceleration, one can claim that time dilation and redshift are ``non-inertial'' phenomena. Moreover, it follows from the considerations in Section \ref{dir-gkk}, that \emi the time of an observer $\Oc$ ``slows down'' in comparison with the time of another sufficiently close observer $\Oc'$, if the four-acceleration of $\Oc$ points out the observer $\Oc'$ and \emii a photon gets redshifted if it moves in the direction of $A$.

Let us emphasize that the connection between the four-acceleration and the phenomena just described, exists in the case of all Killing observers---the metric admitting the observers, does not have to satisfy Einstein equations of any sort. 

As a supplementation of the remarks above, let us recall that (as shown in Section \ref{look}), the four-acceleration $A$ of stationary observers, defined by a Killing vector field $K$, does not change when shifted along integral curves of $K$, and the square $g_{AA}$ is also constant along the curves. In this sense, each stationary observer is {\em uniformly accelerated}.

Let us remark finally that it is a bit unfortunate that both time dilation and redshift perceived by Killing observers, are called ``gravitational''. Since in GR the gravitational field is described by a metric, every mathematical object derived from the metric, can be in principle called ``gravitational''---the meaning of this adjective is very broad, and using it in the names of time dilation and redshift, may wrongly suggest that they are closely related to, e.g., spacetime curvature. Taking into account the conclusions above, it would be perhaps better if the phenomena were called ``accelerational time dilation'' and ``accelerational redshift'', but nowadays it is rather too late to change the terminology.

\section{Summary}

In this paper we considered gravitational time dilation between Killing (stationary) observers and gravitational redshift perceived by them. We proved that there does not exist any measure of strength of the gravitational field, which is independent of the choice of observers, and which guarantees that ``time always slows down as gravity increases''. In particular, no polynomial curvature scalar (like the Kretschmann scalar) can serve as such a measure. We also presented an example, which indicates that the title question cannot be answered in the affirmative, even under the restrictive assumption that Killing observers are static. 

Regarding observer dependent measures of strength of gravity, we did not prove any general result. We considered only one rather natural measure of this sort, that is, the square $g_{AA}$ of the four-acceleration $A$ of Killing observers, and showed that time can ``slow down'' as $g_{AA}$ increases, stays constant or decreases. 

We also presented an interpretation of Killing observers as a GR counterpart of a Newtonian rigid body, and showed that both gravitational time dilation and gravitational redshift perceived by such observers, are not closely related to spacetime curvature, but are conditioned by the four-acceleration of the observers. 

Let us finally address the title question: our answer to it {is ``rather not''. We are unwilling to give a definite negative answer to the question, for the following reason. Namely, we cannot exclude that there exists an observer dependent measure of strength of gravity, which ensures that ``time always slows down as gravity increases''.} But on the other hand, do we really need such a measure? Yet in the case of Killing observers, gravitational time dilation and gravitational redshift, can be strictly described in terms of the scalar field $g_{KK}$ and the four-acceleration $A$ of the {observers. Therefore} we do not need to connect these phenomena with strength of gravity, however it is defined, even for the sake of teaching or popularization.

\paragraph{Acknowledgment} I am very grateful to Peter J. Riggs  for having pointed out some linguistic mistakes in the previous version of the paper.

\appendix

\section{Polar coordinates and gravitational time dilation \label{polar-sec}}

Here we will show that a polar coordinate system on the Euclidean plane, has a lot in common with gravitational time dilation between Killing observers. Perhaps the discussion presented below (stripped of some technicalities), may be helpful in teaching students and amateurs time dilation.

\subsection{A Killing vector field \label{kill-ap}}

The Euclidean plane $\mathbb{E}$ is the space $\R^2$, equipped with a flat Riemannian  metric of the form
\[
g=dx^2+dy^2=dr^2+r^2d\varphi^2,
\]
where $(r,\varphi)$ are polar coordinates related to the Cartesian coordinates $(x,y)$ in the usual way:
\begin{align*}
x&=r\cos\varphi  \quad \quad \quad  {\rm and}  \quad \quad \quad   y=r\sin\varphi.
\end{align*}  

The polar coordinates can be used to define a family of curves on $\mathbb{E}$: each curve is obtained by fixing the coordinate $r$, and is parameterized in a natural way by the coordinate $\varphi$: 
\begin{equation}
\R\ni\varphi\mapsto \big(x(\varphi),y(\varphi)\big)=(r\cos\varphi,r\sin\varphi)\in\mathbb{E}. 
\label{curv}
\end{equation}
Vectors tangent to all these curves form a vector field $K$ on $\mathbb{E}$---the curves are obviously integral curves of $K$. Since $r$ is constant along these curves, and they are naturally parameterized by $\varphi$, the components of $K$ in the polar coordinate system, read $(0,1)$. Thus   
\[
K=\partial_\varphi.
\]

A map $\phi_s$, which belongs to the flow of $K$, maps a point $(x(\varphi),y(\varphi))$ to $(x(\varphi+s),y(\varphi+s))$. This means that $\phi_s$  is a rotation around the point $(x,y)=(0,0)$, through the angle $s$. Thus the flow $\{\phi_s\}$ of $K$ consists of rotations, that is, of maps, which preserve the geometry of the Euclidean plane, and thereby the metric $g$ (since the geometry is defined by the metric). In other words,  $g$ does not change when shifted along the integral curves of $K$, and therefore $K$ is a Killing vector field. Note that the components of the metric $g$ in the polar system, do not depend on $\varphi$, as it should be.

\subsection{``Length dilation''}

\begin{figure}
\psfrag{h}{$\kappa$}
\psfrag{h'}{$\kappa'$}
\psfrag{hz}{$\kappa_{\z_1\z_2}$}
\psfrag{hz'}{$\kappa'_{\z'_1\z'_2}$}
\psfrag{z1}{$\z_1$}
\psfrag{z2}{$\z_2$}
\psfrag{z1'}{$\z'_1$}
\psfrag{z2'}{$\z'_2$}
\psfrag{g1}{$\gamma_1$}
\psfrag{g2}{$\gamma_2$}
\psfrag{o}{$O$}
\psfrag{s}{$s_0$}
\psfrag{x}{$x$}
\psfrag{y}{$y$}
\begin{center}
\includegraphics{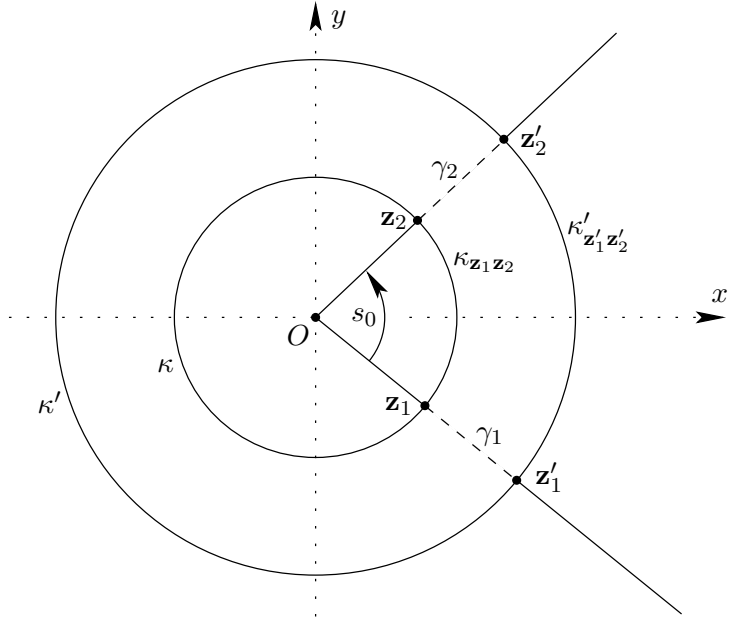}
\end{center}
\caption{``Length dilation'' on the Euclidean plane \label{polar}}
\end{figure}

Denote by $O$ the point $(x,y)=(0,0)$ in the Euclidean plane $\mathbb{E}$, and consider the following curves in $\mathbb{E}$ depicted in Figure \ref{polar}: two circles $\kappa$ and $\kappa'$ with center $O$ and radii $\r$ and $\r'$ satisfying  $\r<\r'$, and two half lines originating at $O$. Obviously, each of the two circles $\kappa,\kappa'$, is a set of constant coordinate $r$, and each of the two half lines, is a set of constant coordinate $\varphi$. The half lines form an angle of $s_0$ radians, and intersect the circles at points $\z_1,\z_2,\z'_1$ and $\z'_2$, determining thereby two arcs $\kappa_{\z_1\z_2}$ and $\kappa'_{\z'_1\z'_2}$. 

The length $\Delta l$ of $\kappa_{\z_1\z_2}$ is obviously $\r s_0$, the length $\Delta l'$ of $\kappa'_{\z'_1\z'_2}$ is $\r' s_0$, and consequently there appears ``length dilation'':
\begin{equation}
D_l=\frac{\Delta l'}{\Delta l}=\frac{\r'}{\r}\neq 1.
\label{l-dil}
\end{equation}
As we will show below, this fact is fully analogous to gravitational time dilation between Killing observers. 

Note first that each of the circles $\kappa$ and $\kappa'$, is an integral curve \eqref{curv} of the Killing vector field $K$. We can then treat the circles as counterparts of world lines of Killing observers. On the other hand, intervals $\gamma_1$ and $\gamma_2$, drawn on Figure \ref{polar} by dashed lines, are geodesics, because they are intervals of the half lines. Thus $\gamma_1$ and $\gamma_2$ are counterparts of the geodesics, used by Killing observers to pair intervals of their world lines. Moreover, the rotation $\phi_{s_0}$ around $O$ through the angle $s_0$ (being an element of the flow of $K$), maps the points $\z_1,\z'_1$ and the interval $\gamma_1$ to, respectively, $\z_2,\z'_2$ and the interval $\gamma_2$. In other words,
\begin{align*}
\z_2&=\phi_{s_0}(\z_1), & \z'_2&=\phi_{s_0}(\z'_1), & \gamma_2&=\phi_{s_0}(\gamma_1),
\end{align*}        
which is in full accordance with the formulas \eqref{zz-Ds}, \eqref{gg-Ds} and \eqref{z'z'-Ds}. The lengths of the intervals $\kappa_{\z_1\z_2}$ and $\kappa'_{\z'_1\z'_2}$ in Equation \eqref{l-dil}, are counterparts of the lapses $\Delta\tau$ and $\Delta\tau'$  of time along intervals of world lines, which appear in Equation \eqref{D-stat}---recall that the lapses are ``spacetime lengths'' of the latter intervals.

Finally, if $\dot{\kappa}$ and $\dot{\kappa}'$ are vectors tangent to the circles $\kappa$ and $\kappa'$, understood as integral curves \eqref{curv} of $K$, then  
\begin{align*}
\r^2&=g(\dot{\kappa},\dot{\kappa})=g(K,K)_{r=\r}\equiv g_{KK}, & \r^{\prime 2}&=g(\dot{\kappa}',\dot{\kappa}')=g(K,K)_{r=\r'}\equiv g'_{KK}.
\end{align*}
Therefore the ``length dilation'' ratio \eqref{l-dil} can be expressed as
\[
D_l=\sqrt{\frac{|g'_{KK}|}{|g_{KK}|}},
\]
that is, exactly as the ratio \eqref{D-stat} of gravitational time dilation.

We thus see that from the point of view of geometry, the example of ``length dilation'' described above, and gravitational time dilation between Killing observers, are the same {phenomenon. There} are some differences between both dilations, but they originate merely from the difference between signatures of the Riemannian metric $g$ on $\mathbb{E}$, and the Lorentzian metrics used {in GR. One} such a difference is that, contrary to \eqref{Agkk}, in the Euclidean plane $A|g_{KK}|\leq 0$, where $A$ is the obvious counterpart of the four-acceleration of Killing observers.   

Let us emphasize finally that the ``length dilation'' described above, appears despite the fact that $\mathbb{E}$ is of zero curvature. 


\end{document}